\documentstyle{mn}
\input{psfig}
\newcommand{\etal}{{et al.~}}

\newcommand{\kmsmpc}{\>{\rm km}\,{\rm s}^{-1}\,{\rm Mpc}^{-1}}
\newcommand{\kms}{\>{\rm km}\,{\rm s}^{-1}}

\newcommand{\Mpc}{\>{\rm Mpc}}

\newcommand{\Msun}{\>{\rm M_{\odot}}}

\newcommand{\mpch}{\>h^{-1}{\rm {Mpc}}}
\newcommand{\yr}{\>{\rm yr}}

\newcommand{\beq}{\begin{equation}}
\newcommand{\eeq}{\end{equation}}

\newcommand{\rmd}{{\rm d}}

\def\gtsima{$\; \buildrel > \over \sim \;$}
\def\ltsima{$\; \buildrel < \over \sim \;$}
\def\prosima{$\; \buildrel \propto \over \sim \;$}
\def\gsim{\lower.7ex\hbox{\gtsima}}
\def\lsim{\lower.7ex\hbox{\ltsima}}
\def\simgt{\lower.7ex\hbox{\gtsima}}
\def\simlt{\lower.7ex\hbox{\ltsima}}
\def\simpr{\lower.7ex\hbox{\prosima}}
\def\la{\lsim}
\def\ga{\gsim}
\def\lta{\la}
\def\gta{\ga}

\newcommand{\apj}{ApJ}
\newcommand{\apjs}{ApJS}
\newcommand{\aj}{AJ}
\newcommand{\mnras}{MNRAS}
\newcommand{\aap}{A\&A}

\newdimen\hssize
\newdimen\hdsize
\begin{document}


\title[Properties of Galaxy Groups in the SDSS]
      {Properties of Galaxy Groups in the SDSS: I.-- The Dependence of
       Colour, Star Formation, and Morphology on Halo Mass}
\author[S.M. Weinmann et al.]
       {Simone M. Weinmann,$^{1,2}$\thanks{E-mail:weinmasi@phys.ethz.ch}
        Frank C. van den Bosch$^{1}$, Xiaohu Yang$^{3}$, H.J. Mo$^{3}$\\
        $^1$Department of Physics, Swiss Federal Institute of
         Technology, ETH H\"onggerberg, CH-8093, Zurich,
         Switzerland\\ 
        $^2$Institute for Theoretical Physics, University of Zurich,
         CH-8057, Zurich, Switzerland\\
        $^3$Department of Astronomy, University of Massachussets, 710
         North Pleasant Street, Amherst MA 01003-9305, USA}


\date{}

\pubyear{2005}

\maketitle

\label{firstpage}


\begin{abstract}
  Using  a large  galaxy group  catalogue constructed  from  the Sloan
  Digital  Sky  Survey  (SDSS  DR2), we  investigate  the  correlation
  between  various galaxy  properties  and halo  mass.   We split  the
  population of galaxies in  early types, late types, and intermediate
  types,  based  on their  colour  and  specific  star formation  rate
  (SSFR).   At fixed  luminosity, the  late (early)  type  fraction of
  galaxies  increases  (decreases) with  decreasing  halo mass.   Most
  importantly, this  mass dependence is  smooth and persists  over the
  entire mass range  probed, without any break or  feature at any mass
  scale. We argue that the  previous claim of a characteristic feature
  on galaxy group scales is  an artefact of the environment estimators
  used.  At  fixed halo  mass, the luminosity  dependence of  the type
  fractions is surprisingly weak, especially over the range $0.25 \lta
  L/L^{*} \lta  2.5$: galaxy type  depends more strongly on  halo mass
  than  on luminosity. In  agreement with  previous studies,  the late
  (early)   type  fraction   increases  (decreases)   with  increasing
  halo-centric radius. However, we find that this radial dependence is
  present  in haloes  of all  masses probed  (down to  $10^{12} h^{-1}
  \Msun$), while  previous studies did not find  any radial dependence
  in haloes with $M \lta  10^{13.5} h^{-1} \Msun$.  We argue that this
  discrepancy owes to the fact  that we have excluded central galaxies
  from our  analysis. We  also find that  the properties  of satellite
  galaxies are strongly correlated with those of their central galaxy.
  In   particular,  the   early   type  fraction   of  satellites   is
  significantly higher  in a  halo with an  early type  central galaxy
  than in a halo  {\it of the same mass} but with  a late type central
  galaxy.  This  phenomenon, which  we call `galactic  conformity', is
  present  in  haloes  of  all   masses  and  for  satellites  of  all
  luminosities.  Finally,  the fraction of  intermediate type galaxies
  is always $\sim 20$  percent, independent of luminosity, independent
  of halo mass, independent of halo-centric radius, and independent of
  whether the  galaxy is a central  galaxy or a  satellite galaxy.  We
  discuss the implications of  all these findings for galaxy formation
  and evolution.
\end{abstract}




\section{Introduction}
\label{sec:intro}

The local  population of galaxies  consists roughly of two  types: red
galaxies, which reveal an early  type morphology and which have little
or  no ongoing  star formation,  and  blue galaxies  with active  star
formation  and  a late-type  morphology.  The  case  for two  distinct
classes of galaxies has recently been strengthened as the use of large
galaxy redshift surveys has shown that the distributions of colour and
star formation rate (SFR) of  the galaxy population are bimodal (e.g.,
Strateva \etal 2001; Blanton  \etal 2003b; Kauffmann \etal 2003, 2004;
Baldry \etal  2004; Brinchmann \etal 2004; Balogh  \etal 2004a,b).  In
addition,  studies  at intermediate  redshifts  have  shown that  this
bimodality exists at least out to $z \simeq 1$ (e.g., Bell \etal 2004;
Tanaka \etal 2005; Weiner \etal 2005), but with different fractions of
galaxies on both sides of the bimodality scale compared to $z=0$ (Bell
\etal 2004; Faber \etal 2005).

An important,  and largely open  question in galaxy  formation regards
the  origin of this  bimodality. In  particular, does  this bimodality
arise  early on (the  `nature' scenario),  or is  it a  consequence of
various  physical  processes that  operate  over  a  Hubble time  (the
`nurture' scenario)?  In particular,  are there two distinct formation
channels,  or are  galaxies being  transformed  from one  type to  the
other? In  the latter case we need  to know where, how  and when these
transformations  occur.   Important   hints  come  from  the  observed
correlations  between galaxy properties  and environment:  galaxies in
dense  environments  (i.e., clusters)  have  predominantly early  type
morphologies (e.g.,  Oemler 1974; Dressler 1980;  Whitmore, Gilmore \&
Jones 1993)  and low  SFRs (e.g., Balogh  \etal 1997,  1999; Poggianti
\etal   1999).   At   first   sight  this   seems   to  suggest   that
cluster-specific  processes, such  as galaxy  harassment  (Moore \etal
1996), ram-pressure stripping (Gunn  \& Gott 1972) and/or interactions
with the  cluster potential  (Byrd \& Valtonen  1990) play  a dominant
role in transforming galaxy morphologies from late to early types, and
in truncating their  SFRs. However, starting with the  work of Postman
\&  Geller  (1984),  it   has  become  clear  that  the  environmental
dependence  of galaxy properties  is not  restricted to  clusters, but
smoothly extends to the scale  of galaxy groups (see also Zabludoff \&
Mulchaey 1998; Tran \etal  2001).  Consequently, it has been suggested
that  group-specific   processes  are  of   paramount  importance  for
transforming  galaxies.  In  particular, the  relatively  low velocity
dispersion  of groups  implies that  galaxy-galaxy merging,  which can
transform  disk  galaxies into  ellipticals  (e.g.,  Toomre \&  Toomre
1972), is effective. In addition, as  soon as a galaxy becomes a group
member, i.e., becomes  a satellite of a bigger  system, it is deprived
of its reservoir of hot gas.  Consequently, it is expected that, after
a delay time in which the galaxy consumes (part of) its cold gas, star
formation in the  galaxy comes to a halt  (Larson, Tinsley \& Caldwell
1980;  Balogh, Navarro \&  Morris 2000).   This process,  often called
strangulation,  provides  a  natural  explanation for  the  increasing
fraction of red galaxies towards denser environments.

Much of the earlier work on the relation between galaxy properties and
environment was  based on incomplete  samples of clusters and  groups. 
With the  advent of large,  homogeneous galaxy surveys, it  has become
possible to investigate  this relation in far more  detail, and over a
much  wider  range of  environments.   In  particular,  using the  Las
Campanas Redshift  Survey (LCRS; Shectman \etal  1996), the Two-Degree
Field  Galaxy Redshift  Survey (2dFGRS;  Colless \etal  2001)  and the
Sloan Digital Sky Survey (SDSS; York \etal 2000; Stoughton \etal 2002)
various authors have investigated the relation between environment and
morphology (e.g., Hashimoto \& Oemler  1999; Goto \etal 2003; Kuehn \&
Ryden  2005),  between  environment  and star  formation  rate  (e.g.,
Hashimoto  \etal  1998;  Lewis  \etal 2002;  Dom\'inguez  \etal  2002;
G\'omez  \etal 2003;  Balogh  \etal 2004a;  Tanaka  \etal 2004;  Kelm,
Focardi  \&  Sorrentino  2005),  and between  environment  and  colour
(e.g.,Tanaka \etal 2004; Balogh \etal 2004b; Hogg \etal 2004).

One of  the numerous results that  have emerged from  these studies is
that  galaxy properties  only seem  to correlate  (significantly) with
environment above  a characteristic surface density,  which is roughly
consistent  with the  characteristic  density at  the  perimeter of  a
cluster or group (Hashimoto \&  Oemler 1999; Lewis \etal 2002; G\'omez
\etal 2003; Goto  \etal 2003; Tanaka \etal 2004;  Balogh \etal 2004a). 
This  has been  interpreted  as further  evidence that  group-specific
processes play a dominant  role in establishing a bimodal distribution
of galaxies (e.g., Postman \& Geller 1984; Zabludoff \& Mulchaey 1998,
2000).  However, it is important to understand the physical meaning of
the density estimators  used.  Most studies parameterize `environment'
through  the  projected  number  density  of galaxies  above  a  given
magnitude  limit.    Typically  this  number   density,  indicated  by
$\Sigma_n$,  is measured  using the  projected distance  to  the $n$th
nearest neighbor, with $n$ typically in the range 5-10 (e.g., Dressler
\etal 1980,  Lewis \etal  2002; G\'omez \etal  2003; Goto  \etal 2003;
Tanaka \etal  2004; Balogh \etal  2004a,b; Kelm \etal  2005). However,
{\it  the  physical  meaning  of  $\Sigma_n$  itself  depends  on  the
  environment}:  in clusters,  where the  number of  galaxies  is much
larger  than $n$, $\Sigma_n$  measures a  {\it local}  number density,
which is a  sub-property of the cluster (i.e.,  $\Sigma_n$ is strongly
correlated  with  cluster-centric  radius).  However,  in  low-density
environments, which  are populated  by haloes which  typically contain
only  one or  two galaxies,  $\Sigma_n$  measures a  much more  global
density, covering a  scale that is much larger than  the halo in which
the galaxy  resides.  This  ambiguous, physical meaning  of $\Sigma_n$
severely   complicates  a   proper  interpretation   of   the  various
correlations  between environment  and galaxy  properties.   Note that
density estimators that use a  fixed metric aperture size, rather than
the distance to  the $n$th nearest neighbor, suffer  from very similar
problems.

Another complication  arises from the fact that  the bimodality scale,
and the fractions of galaxies on either side of it, depend strongly on
luminosity and stellar mass (e.g., Kauffmann \etal 2003; Blanton \etal
2003b; Hogg  \etal 2004;  Baldry \etal 2004;  Kelm \etal  2005).  This
luminosity  dependence  is  also  evident  from a  comparison  of  the
luminosity functions of early and late type galaxies, which shows that
late  (early) types  dominate the  faint (bright)  end  (e.g., Loveday
\etal 1992;  Marzke \& Da Costa  1997; Zucca \etal  1997; Marzke \etal
1998; Blanton \etal  2001; Madgwick \etal 2002).  At  first sight this
seems to  suggest that the morphology  and SFR of a  galaxy is somehow
determined  by its  own  (baryonic)  mass.  On  the  other hand,  this
luminosity/stellar  mass dependence may  also be  a reflection  of the
correlation between the galaxy luminosity function and environment: as
shown by various authors (e.g.,  Hogg \etal 2003; Blanton \etal 2005b;
Mo  \etal   2004;  Hoyle  \etal   2005;  Croton  \etal   2005),  dense
environments  contain  on average  brighter  galaxies.  Therefore,  if
there is a correlation between galaxy properties and environment, this
will introduce a correlation between galaxy properties and luminosity.
Of  course, the  inverse also  holds: a  physical  correlation between
galaxy  properties   and  luminosity  will   introduce  an  observable
correlation between  galaxy properties and  environment. Clearly, when
investigating   the  physical   origin  of   the  bimodality   in  the
distribution of galaxies, it is crucial that one discriminates between
environment dependence and luminosity  dependence in a proper way (see
Girardi  \etal 2003 and  Blanton \etal  2005b for  statistical methods
that address this issue).

\subsection{A physically motivated split of environment}
\label{sec:split}

Within our  current framework for galaxy formation,  in which galaxies
are thought to reside in extended  dark matter haloes, it is useful to
split  the  `environment dependence'  in  three, physically  separate,
components.   Going from  small  to  large scales  these  are (i)  the
dependence on  halo-centric radius, (ii) the dependence  on halo mass,
and (iii) the dependence on  large-scale environment.  In terms of the
halo virial radius, $R_{\rm  vir}$, these effects measure a dependence
on scales $R  < R_{\rm vir}$, $R \simeq R_{\rm vir}$,  and $R > R_{\rm
  vir}$.   Note  that  there  is  a  clear,  physical  motivation  for
considering the  virial radius  as an important  scale: matter  at the
virial radius  has roughly experienced  one dynamical time.   In other
words, a galaxy inside the virial  radius of a given halo can not have
been {\it  dynamically} affected (at  least not significantly)  by any
object that is  located outside of this virial  radius. Thus, if there
is any  galaxy type dependence on  scales $R > R_{\rm  vir}$ this must
arise  from  either  initial  conditions,  or  from  non-gravitational
processes such  as reionization (e.g., Efstathiou  1992) or preheating
(e.g., Mo  \etal 2005).  On  the other hand, most  `nurture' processes
only  introduce a  (radial) dependence  on  scales $R  < R_{\rm  vir}$
Therefore, by investigating  `environment' dependence on scales larger
and  smaller  than the  virial  radius  one may  hope  to  be able  to
determine  which physical  processes  are most  important for  setting
galaxy properties.

Unfortunately, the  presence of a halo mass  dependence may complicate
the situation. Since the  halo mass function is environment dependent,
in that overdense regions contain  on average more massive haloes than
underdense regions (e.g., Lemson \&  Kauffmann 1999; Mo \etal 2004), a
correlation  between galaxy  properties and  halo mass  will  induce a
correlation between galaxy properties and large scale environment. For
example, Mo  \etal (2004) have shown that  the large-scale environment
dependence of  the galaxy luminosity  function of early and  late type
galaxies,  measured on  scales  of  $8 h^{-1}  \Mpc$  by Croton  \etal
(2005), can be  entirely explained as a pure  halo mass dependence. In
addition,  Balogh \etal  (2004a), Blanton  \etal (2004)  and Kauffmann
\etal  (2004) have  shown  that various  galaxy  properties depend  on
environment, even when  the latter is measured over  scales of $\sim 5
h^{-1} \Mpc$, much  larger than the virial radius  of the most massive
clusters. However,  when this large-scale  environmental dependence is
investigated {\it at a fixed small-scale environment}, it is no longer
present (Blanton \etal 2004; Kauffmann \etal 2004, but see also Balogh
\etal  2004a).   Finally,  Goto  \etal  (2003)  have  shown  that  the
morphological  fractions are  constant at  cluster-centric  radii that
exceed  the  virial  radius.   All  these  results  suggest  that  the
environment dependence does not  extent beyond the virial radius. This
is  not only  important because  it  suggests that  processes such  as
reionization and/or preheating have not left a major imprint on galaxy
properties,  but  also because  it  provides  proof  for an  essential
assumption in the halo model  (see Cooray \& Sheth 2002 and references
therein).

A few  studies in the  past have investigated the  correlation between
galaxy  properties   and  halo   mass  using  group   catalogues.   In
particular, Mart\'inez \etal (2002) used a group catalogue constructed
from the  100K data release of  the 2dFGRS by  Merch\'an \& Zandivarez
(2002) and  found that  the  fraction of  early  types decreases  {\it
  continuously}  down to  the lowest  mass  haloes probed  ($M \sim  3
\times 10^{12} \Msun$).  This was confirmed by Yang \etal (2005c), who
used an  independent group  catalogue based on  the completed  2dFGRS. 
Tanaka \etal  (2004) applied the group-finding algorithm  of Huchra \&
Geller (1982) to the first data  release of the SDSS, and examined the
median  SFR  and  morphological  fraction  as function  of  the  group
velocity dispersion $\sigma$.  Splitting the group members into bright
and  faint  galaxies,   they  find  that  neither  the   SFR  nor  the
morphological   fraction  shows   any  significant   correlation  with
$\sigma$, neither for  the bright nor for the  faint galaxies.  Balogh
\etal (2004b) studied the fraction  of red galaxies as function of the
projected density, $\Sigma_5$, and cluster velocity dispersion.  While
they find  a strong dependence  on $\Sigma_5$, for a  fixed luminosity
they find  no dependence  on velocity dispersion  over the  range $300
\kms \lta  \sigma \lta 900  \kms$, corresponding to $3  \times 10^{13}
h^{-1} \Msun \lta M \lta 10^{15}  h^{-1} \Msun$ (cf., De Propris et al
2004;   Goto   2005).    Although   the   comparison   is   far   from
straightforward,  these findings  of  Tanaka \etal  (2004) and  Balogh
\etal  (2004b) seem difficult  to reconcile  with those  of Mart\'inez
\etal (2002)and  Yang et al  (2005c).  A more  in-depth investigation,
based on a large and well  defined sample is required in order to shed
some  light on these  issues, and  to examine  any possible  halo mass
dependence in more detail.

\subsection{The purpose of this paper}
\label{sec:purpose}

In  this  paper  we  investigate  the  dependence  of  various  galaxy
properties, in particular colour, SFR, and concentration, on halo mass
and halo-centric  radius.  To  that extent we  construct a  SDSS group
catalogue using  the halo-based  group finder of  Yang \etal  (2005a). 
This group finder  has been well tested, and  yields high completeness
and a low  fraction of interlopers. Halo masses  are assigned based on
the group  luminosity, which,  as we will  show, yields  more reliable
mass estimates than the  conventional velocity dispersion of the group
members.

We  use the  resulting group  catalogue  to examine  the fractions  of
various  galaxy  types  as  function  of luminosity,  halo  mass,  and
halo-centric radius.   Since haloes of different  masses host galaxies
of different luminosities  (e.g., Yang, Mo \& van  den Bosch 2003; van
den Bosch,  Yang \& Mo 2003),  it is important  to separate luminosity
dependence from halo mass dependence.  We address this by studying the
halo mass dependence at fixed luminosity and vice versa.

This  paper is  organized as  follows. In  Section~\ref{sec:class}, we
describe the data and our classification of galaxy types based on both
colour  and SFR.   In Section~\ref{sec:groupcat}  we present  our SDSS
group catalogue, which we  use in Section~\ref{sec:res} to investigate
the   relation  between   galaxy  properties   and  halo   mass.   The
implications  of  our findings  for  the  formation  and evolution  of
galaxies  is discussed in  Section~\ref{sec:disc}, while  we summarize
our results  in Section~\ref{sec:concl}.  The paper  also contains two
appendices:  Appendix~A  gives a  detailed  description  of our  group
finder and Appendix~B presents a  number of tests based on mock galaxy
redshift surveys to illustrate the robustness of our results.

When  required  we  adopt   a  standard  $\Lambda$CDM  cosmology  with
$\Omega_m=0.3$ and $\Omega_{\Lambda} =  0.7$. Units that depend on the
Hubble constant are expressed in terms of $h \equiv (H_0/100\kmsmpc)$.

\section{Classifying Galaxies based on colour and star formation rate}
\label{sec:class}

\subsection{The data}
\label{sec:data}

The data used in this paper is taken from the Sloan Digital Sky Survey
(SDSS; York \etal  2000), a joint, five passband  ($u$, $g$, $r$, $i$,
$z$)  imaging  and  medium-resolution  ($R \sim  1800$)  spectroscopic
survey.   In particular, we  use the  New York  University Value-Added
Galaxy  Catalogue  (NYU-VAGC), which  is  described  in Blanton  \etal
(2005a). The NYU-VAGC  is based on the SDSS  Data Release 2 (Abazajian
\etal  2004), but with  an independent  set of  significantly improved
reductions.  From  this catalogue we  select all galaxies in  the Main
Galaxy Sample,  i.e., galaxies  with an extinction  corrected apparent
magnitude  brighter  than  $r=18$.   We  prune this  sample  to  those
galaxies in  the redshift  range $0.01  \leq z \leq  0.20$ and  with a
redshift completeness $c > 0.7$.  This leaves a grand total of 184,425
galaxies with a sky coverage of $\sim 1950$ ${\rm deg}^2$.

In addition to these data, we also use estimates of the stellar masses
and the star formation rates (SFRs) obtained by Kauffmann \etal (2003)
and  Brinchmann  \etal   (2004),  respectively.   Stellar  masses  are
obtained from  the strength  of the $4000${\AA}  break and  the Balmer
absorption-line  index H$\delta_{A}$ as  described in  Kauffmann \etal
(2003), while the SFR is  obtained using various emission lines in the
SDSS spectra as described in Brinchmann \etal (2004). In this paper we
mainly use the specific  star formation rate (hereafter SSFR), defined
as the ratio of the SFR  (in $\Msun \yr^{-1}$) to the stellar mass (in
$\Msun$). The SSFRs used are the average values of the full likelihood
distributions  obtained by  Brinchmann et  al.  The  NYU-VAGC  and the
stellar    mass    and     SFR    catalogues    are    all    publicly
available\footnote{The       NYU-VAGC       is      available       at
  http://wassup.physics.nyu.edu/vagc/\#download,  while the catalogues
  with   stellar   masses   and    SFRs   can   be   downloaded   from
  http://www.mpa-garching.mpg.de/SDSS/}.    We   have  matched   these
catalogues yielding  a (dust-corrected)  stellar mass and  current SFR
(corrected  for fiber aperture)  for 179,197  of the  184,425 galaxies
($\sim 97$ percent) in our sample.

Throughout this  paper we use the Petrosian  magnitudes, corrected for
Galactic  extinction using the  dust maps  of Schlegel,  Finkbeiner \&
Davis (1998).  In order to minimize the errors due to uncertainties in
the  k-correction we follow  Blanton \etal  (2005a) and  k-correct all
magnitudes to  a redshift of  $z=0.1$ using the Blanton  \etal (2003c)
model. We use  the notation $^{0.1}M_r$ and $^{0.1}r$  to indicate the
resulting   absolute  and   apparent  magnitudes   in   the  $r$-band,
respectively.

The   spectroscopic  survey  of   the  SDSS   suffers  from   a  small
incompleteness due  to (i) fiber collisions (6  percent), (ii) spectra
that did not  allow for a useful determination of  the redshift ($< 1$
percent),  and (iii) galaxies  that were  too close  to a  bright star
(Blanton \etal 2004).  Of these, the fiber collision incompleteness is
the   most   important  one,   especially   because   it  creates   an
incompleteness which  is correlated with  the local number  density of
galaxies. Since  in this paper we  are not interested  in any absolute
number  densities, we  have not  attempted to  correct the  survey for
these fiber collisions.   Our main focus is on  the {\it fractions} of
various galaxy  types {\it as  a function of environment}.   Since the
galaxies missed because of fiber collisions are a purely random subset
of  the galaxies in  the target  field, their  absence should  have no
impact on the type fractions discussed here.

\begin{figure*}
\centerline{\psfig{figure=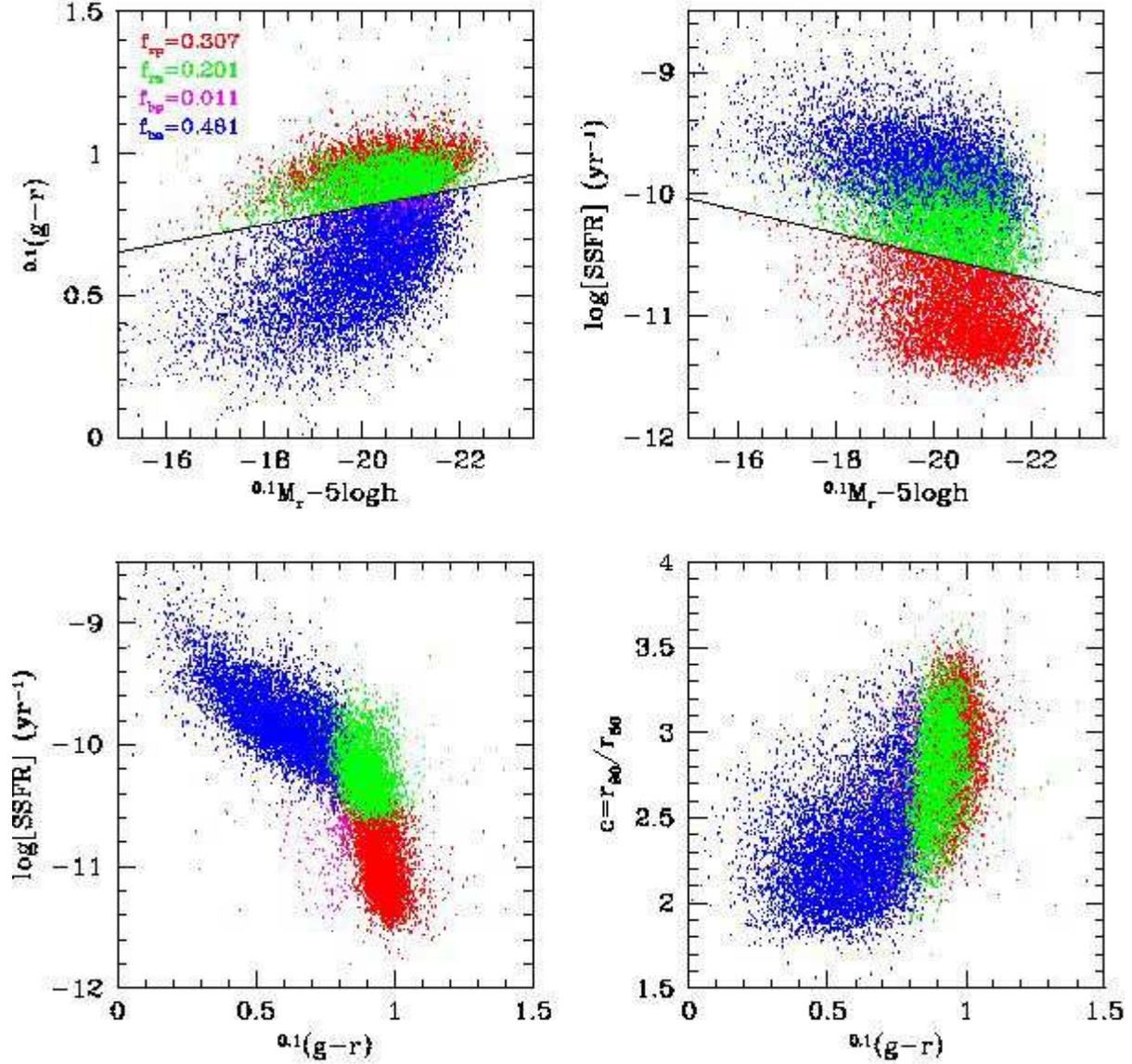,width=0.9\hdsize}}
\caption{The upper left-hand panel shows the colour-magnitude relation 
  for  the galaxies  in our  sample.   The solid  line corresponds  to
  equation~(\ref{colmagcrit}) and  splits the galaxies  into `red' and
  `blue'   subsamples.    The  upper   right-hand   panel  shows   the
  SSFR-magnitude  relation for  the  same galaxies.   The solid  lines
  corresponds to equation~(\ref{ssfrmagcrit})  and splits the galaxies
  into  `active and `passive'  subsamples.  Galaxies  are colour-coded
  according to  these classifications: red  dots (30.7 percent  of the
  population; hereafter  `early'-types) are `red'  and `passive', blue
  dots (48.1  percent of  the population; hereafter  `late'-types) are
  `blue'  and `active', green  dots (20.1  percent of  the population;
  hereafter `intermediate'-types) are  `red' and `active', and magenta
  dots are `blue' and `passive'. Since the latter only make up for 1.1
  percent of all galaxies we do  not consider them any further in this
  paper.   The lower  left-hand panel  plots the  SSFR as  function of
  colour.   Note  how  the  intermediate  types  are  located  at  the
  cross-section  of the early  and late  type branches.   Finally, the
  lower  right-hand  panel plots  the  concentration  of each  galaxy,
  defined as the ratio of $r_{90}$ to $r_{50}$, as function of colour.
  For clarity, only  a random subsample of 10  percent of all galaxies
  is shown.}
\label{fig:galprop}
\end{figure*}

\subsection{Defining galaxy types}
\label{sec:typedef}

The  main purpose  of this  paper is  to investigate  how  galaxy type
correlates with halo mass.  Roughly, the galaxy population consists of
two types:  `early types', which  have red colours, low  specific star
formation  rates, and are  morphologically reminiscent  of ellipticals
and S0s,  and `late types',  which have blue colours,  relatively high
specific star  formation rates, and are  morphologically classified as
spiral galaxies.

Unfortunately, whether a galaxy is  termed `early' or `late' is fairly
subjective, and many different approaches  have been used in the past,
including morphological quantifiers (e.g.  Tran \etal 2001; Goto \etal
2003),  star  formation  rate  indicators  (e.g.,  Lewis  \etal  2002;
Mart\'{i}nez \etal 2002; Dom\'{i}nguez \etal 2002; Balogh \etal 2004a;
Tanaka \etal 2005) and broad-band colours (Strateva \etal 2001; Baldry
\etal 2004; Balogh  \etal 2004b; Goto \etal 2004).  The 2dFGRS and the
SDSS have  clearly revealed  that the distributions  of many  of these
parameters are  (to some extent)  bimodal (e.g., Strateva  \etal 2001;
Madgwick \etal  2002; Blanton \etal  2003b).  Although this  makes the
split   more   objective,    the   non-uniqueness   of   the   various
type-classifications creates some  ambiguity.  For example, a genuine,
star-forming  disk galaxy  may  appear red  due  to strong  extinction
(e.g., when  seen edge-on), and thus  be termed `early  type' based on
its colour, while the SFR and morphology quantifiers would classify it
as a `late-type'.

To partially  sidestep these  difficulties we classify  galaxies using
{\it  both}  colour  and  specific  star formation  rate.   The  upper
left-hand panel  of Fig.~\ref{fig:galprop} shows  the colour-magnitude
(CM) relation  for a random subsample  of 10 percent of  all galaxies. 
The CM  relation is clearly  bimodal, revealing a narrow  red sequence
and a much broader blue sequence (see also Blanton \etal 2003b; Baldry
\etal 2004; Hogg  \etal 2004; Bell \etal 2004).   The thick solid line
corresponds to
\begin{equation}
\label{colmagcrit}
^{0.1}(g-r) = 0.7 - 0.032 \left(^{0.1}M_r - 5 {\rm log} h + 16.5\right)
\end{equation}
with  $^{0.1}M_r$  the  absolute   magnitude  in  the  SDSS  $r$-band,
k-corrected to  $z=0.1$.  We term  galaxies that fall above  this line
`red', and galaxies below this line as `blue'.

\begin{figure*}
\centerline{\psfig{figure=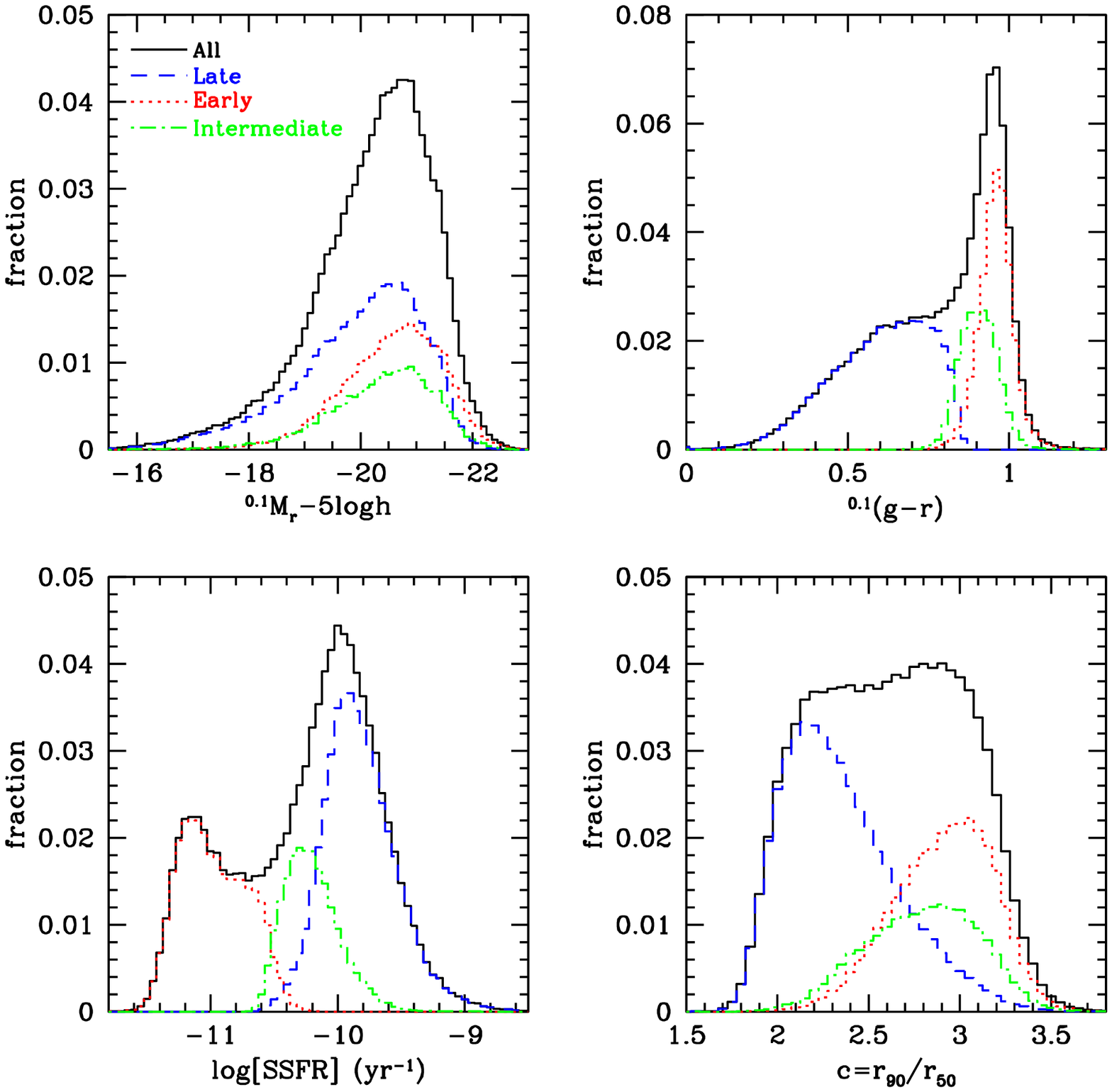,width=0.9\hdsize}}
\caption{Histograms of the distribution of galaxies as function of
  absolute   $r$-band  magnitude   (upper-left   panel),  $g-r$-colour
  (upper-right  panel),  specific   star  formation  rate  (lower-left
  panel), and  concentration (lower-right  panel). In addition  to the
  distributions for the full  sample of galaxies (black, solid lines),
  we also show the distributions  for late types (blue, dashed lines),
  early  types (red,  dotted  lines), and  intermediate types  (green,
  dot-dashed lines). Note that the intermediate types have colours and
  SSFRs that are  intermediate to those of early-  and late types, but
  have luminosities  and concentrations that are  reminiscent to those
  of the early types.}
\label{fig:prophis}
\end{figure*}

The upper right-hand panel of Fig.~\ref{fig:galprop} plots the SSFR as
function  of absolute  magnitude.   Similar to  the  CM relation,  the
distribution  is clearly  bimodal (see  also  Fig.~\ref{fig:prophis}). 
The thick solid line corresponds to
\begin{equation}
\label{ssfrmagcrit}
{\rm log}({\rm SSFR}) = -10.0 + 0.094 \left(^{0.1}M_r - 5 {\rm log} h + 
15.0\right)
\end{equation}
and  roughly  describes the  magnitude  dependence  of the  bimodality
scale.  Galaxies  that fall above  this line are termed  `active', and
those below it `passive'.

Galaxies that  are `red'  and `passive' are  indicated by red  dots in
Fig.~\ref{fig:galprop}  and  make  up   30.7  percent  of  the  entire
population. In what follows we refer to these as early types. Galaxies
that are  `blue' and  `active' are represented  by blue dots,  make up
48.1 percent of  the population, and will hereafter  be referred to as
late-types. A fraction  of 20.1 percent of all  galaxies are `red' and
`active'. These  are represented by  green dots and will  hereafter be
referred to as intermediate types.  The final class of galaxies, those
that are  both `blue' and `passive',  only make up 1.1  percent of all
galaxies (magenta dots  in Fig.~\ref{fig:galprop}). Thus, although our
classification allows  for four classes, in practice,  98.9 percent of
all  galaxies belongs  to only  three  of these.   This suggests  that
galaxies  occupy  only  a   restricted  subspace  of  the  colour-SSFR
parameter space.   Indeed, as  shown in the  lower left-hand  panel of
Fig.~\ref{fig:galprop},  galaxies  follow  a  roughly  one-dimensional
distribution in this plane.  Most importantly, the different types are
clearly separated, with the intermediate types occupying the region in
between  the  early  and  late  types  (hence  our  choice  for  their
nomenclature).   The  clarity  with  which the  various  galaxy  types
separate  out in this  colour-SSFR sequence  gives a  strong, physical
motivation for our classification  scheme.  Note that the intermediate
types seem to occupy the region where the late and early type branches
overlap. This  suggests that they consist  of a mix of  early and late
types, rather than constitute a physically separate class.

Most of  the `blue' and  `passive' galaxies (magenta points)  fall off
the  colour-SSFR sequence:  they are  clearly  not part  of the  major
population of galaxies.  Because of  this, and since they only make up
a negligible fraction  of the total population, we  no longer consider
them in this paper.

The  lower  right-hand   panel  of  Fig.~\ref{fig:galprop}  plots  the
concentration  parameter  $c  \equiv  r_{90}/r_{50}$  as  function  of
colour.  Here $r_{90}$ and $r_{50}$  are the radii that contain 90 and
50 percent of the  Petrosian $r$-band flux, respectively. As expected,
early types  are,   on  average,  more   centrally  concentrated  than
late types. Note also that the intermediate types cover the full range
of concentrations  expected given their  colour. In other  words, they
are  not predominantly  low or  high concentration  systems.   

Figure~\ref{fig:prophis}  shows  histograms  of the  distributions  of
absolute  magnitude, $^{0.1}(g-r)$-colour,  log(SSFR),  and $c$.   The
dashed,  dotted and dot-dashed  curves show  the contributions  due to
late,  early,  and intermediate  types,  respectively.   Note that  no
correction  has been applied  for Malmquist  bias (i.e.,  no $1/V_{\rm
  max}$ weighting  has been applied), so that  the distributions shown
do not reflect true number density distributions: they merely serve as
an  illustration.  Note  how  the  early and  late  types are  clearly
separated in terms of colour  and SSFR (by construction), and that the
intermediate types  have distributions that are  truly intermediate to
those of the early and late types.  The $c$-distributions of early and
late types are clearly skewed towards the opposite extremes, but still
show a large range of overlap.  Although the intermediate types have a
$c$-distribution that is more reminiscent  of that of the early types,
they have the same $c$-distribution  as late-type galaxies of the same
colour (cf., lower right-hand panel of Fig.~\ref{fig:galprop})

Our class of early types thus  consists of red galaxies with a passive
SFR and  a high concentration,  consistent with a typical  elliptical. 
Our  class of  late  types consists  of  galaxies that  are blue,  are
actively forming  stars, and  have low concentrations,  all consistent
with a typical  spiral galaxy.  The nature of  our intermediate types,
however, is less  clear. They are defined as  galaxies that are `red',
yet `active'.  Therefore,  it is tempting to interpret  them as dusty,
star forming  galaxies. One possibility is  that they are,  to a large
extent, made up of edge-on  disk galaxies where the orientation causes
an  enhanced extinction. On  the other  hand, Brinchmann  \etal (2004)
have  stated that  due to  degeneracies between  age,  metallicity and
dust, the SFR  cannot be constrained better than to a  factor of 10 at
colours  redder than  $^{0.1}(g-r)=0.7$.  Therefore,  the intermediate
types may also  consist of early type galaxies for  which the SSFR has
been  overestimated.  Most  likely,  our class  of intermediate  types
contains examples of  both. Indeed, as we will  show below, their halo
occupation statistics strongly  suggest that they consist of  a mix of
both early and late types.

\section{The SDSS Group Catalogue}
\label{sec:groupcat}

\subsection{The group finding algorithm}
\label{sec:groupfinder}

In order to study the relation  between galaxy types and halo mass, we
construct  a   group  catalogue  from  the  SDSS   data  described  in
Section~\ref{sec:data},  using all galaxies  in our  sample, including
those for which no stellar mass or SSFR is available.

Our working definition  of a galaxy group is  the ensemble of galaxies
that reside in the same  dark matter parent halo; galaxies that reside
in subhaloes  are considered  to be group  members that belong  to the
parent halo  in which the subhalo  is located.  The  properties of the
halo  population   in  the   standard  $\Lambda$CDM  model   are  well
understood, largely  due to a combination of  $N$-body simulations and
analytical models.  Recently, Yang  \etal (2005a, hereafter YMBJ) used
this  knowledge  to develop  a  new  group-finding  algorithm that  is
optimized  to group  galaxies according  to their  common  dark matter
halo, and which  has been thoroughly tested with  mock galaxy redshift
surveys.   In brief,  the method  works as  follows.   First potential
group  centers   are  identified  using   a  Friends-Of-Friends  (FOF)
algorithm or an isolation criterion.  Next, the total group luminosity
is estimated  which is converted into  an estimate for  the group mass
using  an assumed mass-to-light  ratio. From  this mass  estimate, the
radius and  velocity dispersion of the corresponding  dark matter halo
are estimated  using the virial equations,  which in turn  are used to
select group members in redshift  space. This method is iterated until
group memberships  converge. A more  detailed description is  given in
Appendix~A.  The  basic idea  behind this group  finder is  similar to
that of the matched filter algorithm developed by Postman \etal (1996)
(see also Kepner  \etal 1999; White \& Kochanek  2002; Kim \etal 2002;
Kochanek \etal  2003; van  den Bosch \etal  2004, 2005a),  although it
also makes use of the galaxy kinematics.

In YMBJ the performance of this  group finder has been tested in terms
of  completeness of  true  members and  contamination by  interlopers,
using detailed mock galaxy redshift surveys.  The average completeness
of  individual groups is  $\sim 90$  percent and  with only  $\sim 20$
percent  interlopers.  Furthermore, the  resulting group  catalogue is
insensitive  to  the initial  assumption  regarding the  mass-to-light
ratios, and the group finder  is more successful than the conventional
FOF method  (e.g., Huchra  \& Geller 1982;  Ramella, Geller  \& Huchra
1989; Merch\'an  \& Zandivarez 2002;  Eke \etal 2004a)  in associating
galaxies according to their common dark matter haloes.

Thus far this group finder has  been applied to the 2dFGRS (Yang \etal
2005a) and used to study  the two-point correlation function of groups
(Yang \etal  2005b), the galaxy  occupation statistics of  dark matter
haloes  (Yang \etal  2005c), the  phase-space parameters  of brightest
halo galaxies  (van den Bosch \etal 2005b),  and the cross-correlation
between galaxies and groups (Yang \etal 2005d). In this paper we apply
it to the  SDSS. The resulting group catalogue  is used to investigate
the relation between various galaxy properties and halo mass.

\subsection{Estimating group masses}
\label{sec:groupmass}

In order to infer halo occupation statistics from our group samples it
is crucial  that we can estimate  the halo masses  associated with our
groups. For individual, rich  clusters one could in principle estimate
halo masses using the kinematics of the member galaxies, gravitational
lensing of background sources, or the temperature profile of the X-ray
emitting gas.   For most groups,  however, no X-ray emission  has been
detected, and  no lensing  data is available.   In addition,  the vast
majority  of the  groups in  our sample  contain only  a  few members,
making  a  dynamical mass  estimate  based  on  its members  extremely
unreliable  (see Appendix  B).   We  thus need  to  adopt a  different
approach to  estimate halo  masses.  Following YMBJ  we use  the group
luminosity to assign masses to  our groups. The motivation behind this
is  that one  naturally expects  the group  luminosity to  be strongly
correlated with halo  mass (albeit with a certain  amount of scatter). 
Since  the group  luminosity is  dominated by  the  brightest members,
which are  exactly the  ones that  can be observed  in a  flux limited
survey  like  the  SDSS,   the  determination  of  the  (total)  group
luminosity  is far  more  robust  than that  of  the group's  velocity
dispersion when the number of group members is small.

\begin{figure*}
\centerline{\psfig{figure=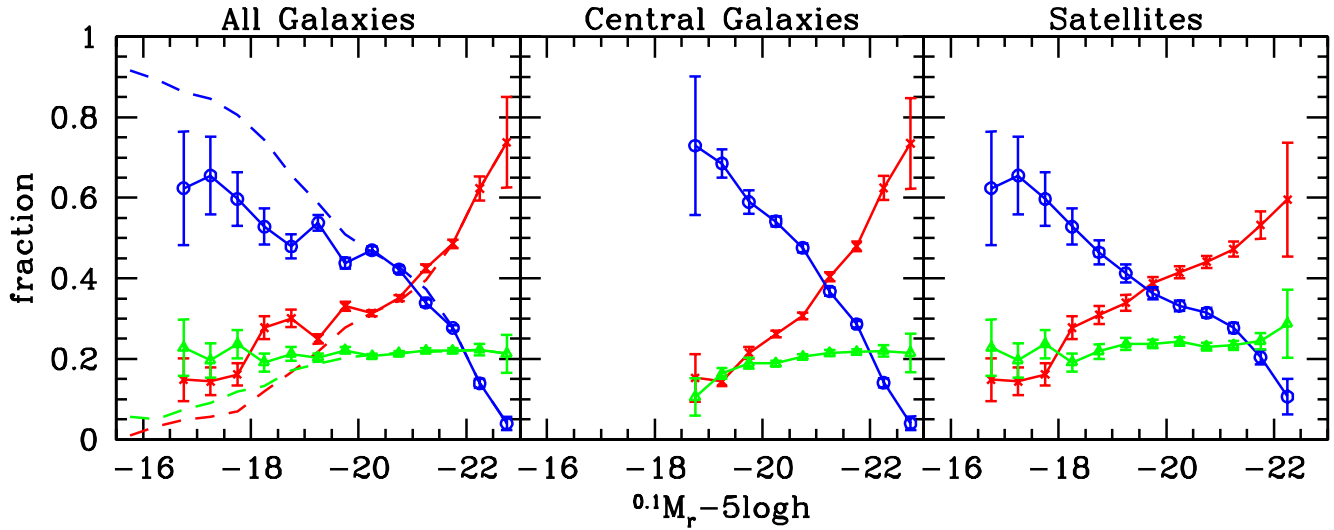,width=\hdsize}}
\caption{The fraction of late types (open circles with blue lines),
  early types  (crosses with red  lines) and intermediate  types (open
  triangles with  green lines) as  function of absolute  magnitude (in
  $r$-band,  k-corrected  to  $z=0.1$).   Results are  shown  for  all
  galaxies in groups (left-hand  panel), the central (brightest) group
  galaxies (middle panel), and  satellite galaxies (right-hand panel). 
  In addition, the  dashed lines in the left-hand  panel show the type
  fractions for all galaxies, including those not assigned to a group.
  Errorbars  indicate Poisson  errors.  The  fraction of  late (early)
  types  decreases (increases)  strongly with  increasing  luminosity. 
  Note that  the luminosity  dependence is significantly  stronger for
  central  galaxies  than  for  satellite  galaxies. See  text  for  a
  detailed discussion.}
\label{fig:lumdep}
\end{figure*}

Clearly, because of  the flux limit of the  SDSS, two identical groups
observed at different redshifts will have a different $L_{\rm group}$,
defined as  the summed luminosity  of all its identified  members.  To
circumvent this bias we first  need to bring the group luminosities to
a  common scale.   One possibility  is to  use the  {\it  total} group
luminosity, $L_{\rm total}$, which one might define according to
\begin{equation}
\label{Ltotal}
L_{\rm total} = L_{\rm group} \frac{\int_0^{\infty} \Phi(L) \, L \, dL}
{\int_{L_{\rm lim}}^{\infty} \Phi(L) \, L \, dL}\,.
\end{equation}
Here $L_{\rm lim}$  is the minimum luminosity of a  galaxy that can be
observed at  the redshift  of the group,  and $\Phi(L)$ is  the galaxy
luminosity function in the $^{0.1}r$-band.  Although this approach has
been used  by many earlier  analyzes (e.g., Tucker 2000;  Merch\'an \&
Zandivarez 2002; Kochanek \etal 2003; Eke \etal 2004b), it is based on
the assumption  that the galaxy  luminosity function in groups  is the
same as that of field galaxies,  independent of the mass of the group. 
It  has  been shown,  however,  that  the  galaxy luminosity  function
depends on both halo mass  and environment (Yang \etal 2003, 2005c; Mo
\etal  2004;   Zheng  \etal  2004;   Croton  \etal  2005;   Cooray  \&
Milosavljevi\'c  2005).   Therefore we  follow  YMBJ  and  use a  more
empirical approach.  A nearby  group selected in an apparent magnitude
limited  survey should  contain all  of its  members down  to  a faint
luminosity.  We can therefore use these nearby groups to determine the
relation  between the  group luminosity  obtained using  only galaxies
above a bright luminosity limit and that obtained using galaxies above
a  fainter   luminosity  limit.    Assuming  that  this   relation  is
redshift-independent,  one can  correct the  luminosity of  a high-$z$
group,  where   only  the  brightest  members  are   observed,  to  an
empirically normalized luminosity scale.

As  common  luminosity  scale   we  use  $L_{19.5}$,  defined  as  the
luminosity of  all group  members brighter than  $^{0.1}M_r =  -19.5 +
5\log  h$.  To  calibrate  the relation  between  $L_{\rm group}$  and
$L_{19.5}$  we first  select  all  groups with  $z  \leq 0.09$,  which
corresponds to the redshift for which a galaxy with $^{0.1}M_r = -19.5
+ 5\log  h$ has an apparent  magnitude that is equal  to the magnitude
limit of the  survey.  For groups with $z > 0.09$  we use this `local'
calibration  between $L_{\rm  group}$ and  $L_{19.5}$ to  estimate the
latter.  Detailed tests (see YMBJ) have shown that the resulting group
luminosities are significantly more reliable than $L_{\rm total}$.

The final step is to obtain  an estimate of the group (halo) mass from
$L_{19.5}$.   This  is done  using  the  assumption  that there  is  a
one-to-one relation between $L_{19.5}$  and halo mass.  For each group
we determine  the number density of  all groups brighter  (in terms of
$L_{19.5}$)  than the  group in  consideration.  Using  the  halo mass
function  corresponding to a  $\Lambda$CDM concordance  cosmology with
$\Omega_m=0.3$,  $\Omega_{\Lambda}=0.7$, $h=H_0/(100  \kmsmpc)  = 0.7$
and $\sigma_8=0.9$  we then find the  mass for which  the more massive
haloes have the  same number density.  Although this  has the downside
that it  depends on  cosmology, it is  straightforward to  convert the
masses derived here to any  other cosmology. An obvious shortcoming of
this  method is  that the  true  relation between  $L_{19.5}$ and  $M$
contains  some scatter.   This scatter  will result  in errors  in the
inferred halo mass.   However, as long as the  scatter is sufficiently
small, which we believe to be  the case, given, for example, the small
observed  scatter  in  the   Tully-Fisher  relation,  this  method  of
assigning group  masses is expected to be  significantly more accurate
than using the velocity dispersion of group members.  In Appendix~B we
use detailed mock galaxy redshift  surveys to demonstrate that this is
indeed the case (see also YMBJ and Yang \etal 2005c).

Finally we note  that not all groups can have a  halo mass assigned to
them.  First of all, the  mass estimator described above does not work
for groups in which all members  are fainter than $^{0.1}M_r = -19.5 +
5\log h$. Secondly, the combination  of $L_{19.5}$ and redshift may be
such that we  know that the halo catalogue  is incomplete, which means
that there is a significant number of groups at this redshift with the
same $L_{19.5}$ but for which the individual galaxies are too faint to
be detected.  Since our mass  assignment is based on the assumption of
completeness, any group beyond the completeness redshift corresponding
to its  $L_{19.5}$ is not assigned  a halo mass (see  Yang \etal 2005a
for details).

\subsection{The SDSS group catalogue}
\label{sec:catalogue}

Applying our group finder to  the sample of SDSS galaxies described in
Section~\ref{sec:data} yields a group catalogue of 53,229 systems with
an estimated  mass. These groups contain  a total of  92,315 galaxies. 
The  majority of  the groups  (37,216 systems)  contain only  a single
member, while there are 9220 binary systems, 3073 triplet systems, and
3720 systems  with four members or  more. In what follows  we refer to
the brightest galaxy in each  group as the `central' galaxy, while all
others are termed `satellites'.

This    SDSS    group    catalogue    is   publicly    available    at
http://www.astro.umass.edu/$^{\sim}$xhyang/Group.html    \footnote{This
  website also contains our 2dFGRS group catalogue as well as detailed
  mock galaxy redshift surveys}.   For each group-member the catalogue
contains magnitudes in the five  SDSS bands ($u$, $g$, $r$, $i$, $z$),
Petrosian  radii,  a velocity  dispersion,  and,  for 89,232  galaxies
($\sim 97$ percent of all  group members) the stellar mass and present
day  SFR.   In  addition  to  group memberships,  the  catalogue  also
contains   estimates  of   the   group's  characteristic   luminosity,
$L_{19.5}$, and its mass (derived using the method described above).

\begin{figure*}
\centerline{\psfig{figure=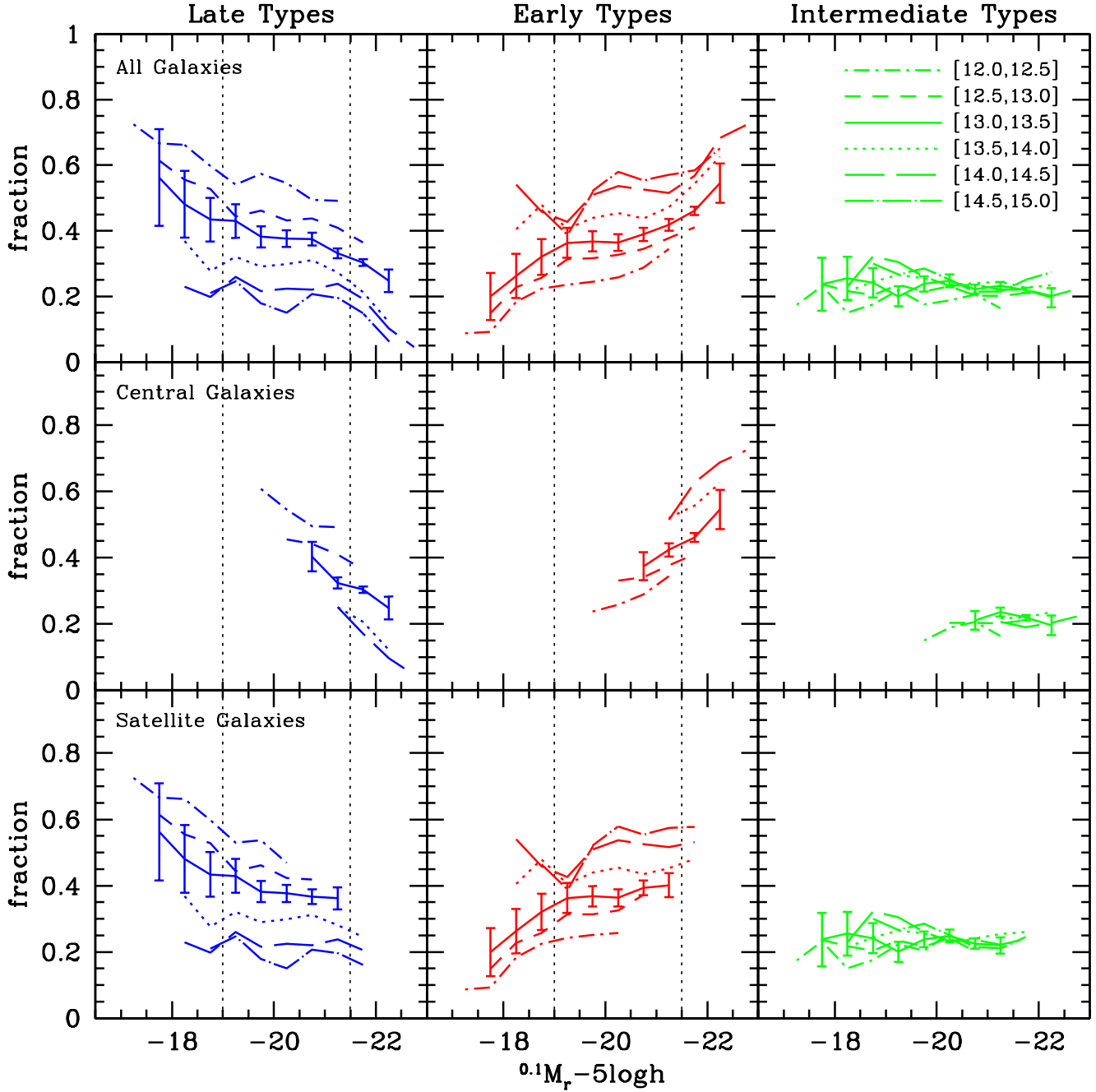,width=\hdsize}}
\caption{The fractions of late types (left-hand panels), early types
  (middle  panels)  and  intermediate  types  (right-hand  panels)  as
  function  of their  absolute magnitude.  Results are  shown  for all
  galaxies (upper panels), central galaxies (middle row of panels) and
  satellite galaxies (lower panels), and  for 6 different mass bins as
  indicated (the values in square brackets indicate the range of ${\rm
    log}(M/h^{-1}\Msun)$).  Results are only shown for mass-luminosity
  bins that  contain at  least 50 galaxies  in total, and  for clarity
  (Poissonian) errorbars are  only shown for one mass  bin.  Note that
  the fraction of early and late types at fixed luminosity is strongly
  mass-dependent, while  luminosity dependence  at fixed mass  is only
  evident at the bright and faint ends. In the intermediate range $-19
  \geq  ^{0.1}M_r -  5  {\rm  log}h \geq  -21$  (indicated by  dotted,
  vertical lines), the luminosity dependence is surprisingly weak, for
  all halo  masses.  Note that  the fraction of intermediate  types is
  completely independent  of both luminosity  and halo mass,  and does
  not  depend  on  whether  the  galaxy  is  a  central  galaxy  or  a
  satellite.}
\label{fig:typelum}
\end{figure*}

\section{Results}
\label{sec:res}

Using the SDSS group catalogue  described above, and the definition of
galaxy   types   discussed   in  Section~\ref{sec:typedef},   we   now
investigate the ecology of galaxies.

\subsection{Dependence on Luminosity}
\label{sec:lumdep}

We start  our investigation  by computing how  galaxy type  depends on
luminosity.   The left-hand panel  of Fig.~\ref{fig:lumdep}  plots the
various type  fractions as function  of the absolute magnitude  in the
$^{0.1}r$-band.  For  each luminosity  bin, we only  consider galaxies
with  $0.01 \leq  z  \leq z_{\rm  max}$,  where $z_{\rm  max}$ is  the
redshift out to which a galaxy  at the faint end of the luminosity bin
has an apparent magnitude that is  equal to the flux limit of the SDSS
($r = 17.77$).  In other words, each magnitude bin is a volume limited
sample.   The  points  connected  by  the  solid  lines  indicate  the
fractions of all galaxies that are  member of a group with an assigned
mass. Results are only shown for luminosity bins that contain at least
50  galaxies  in total  and  errorbars  are  calculated using  Poisson
statistics.

As is well known, the late (early) type fraction decreases (increases)
strongly with  increasing luminosity (e.g., Baldry  \etal 2004; Balogh
\etal  2004b; Kelm  \etal 2005).   The fraction  of  intermediate type
systems,  however,  is  remarkably  constant  at  $\sim  20$  percent,
virtually independent of luminosity.

The dashed lines  indicate the type fractions when  {\it all} galaxies
are  considered, including  those that  have  not been  assigned to  a
group.  Note  that these fractions differ substantially  from those of
the group members  at the faint end. This is a  first indication for a
mass dependence  of the type  fractions; since our group  catalogue is
incomplete at the low mass end,  because its members are too faint for
a mass estimate  (see Section~\ref{sec:groupmass}), the faint galaxies
that do make it into the group catalogue are mainly satellite galaxies
in more  massive haloes.   The results shown  here suggest  that these
have a  lower late type fraction  than galaxies of  the same magnitude
but which reside in more massive haloes.

The  middle and  right-hand panels  of Fig.~\ref{fig:lumdep}  plot the
type fractions for central and satellite galaxies, respectively (again
using only galaxies in groups  with an assigned halo mass). This shows
that the luminosity  dependence of the type fractions  is stronger for
central  galaxies than  for satellite  galaxies. A  similar  trend was
previously noted by Yang \etal  (2005c) from an analysis of the 2dFGRS
group  catalogue.  Note that  the fraction  of intermediate  types is,
within the errors, equally large among central and satellite galaxies,
independent of luminosity.

\begin{figure*}
\centerline{\psfig{figure=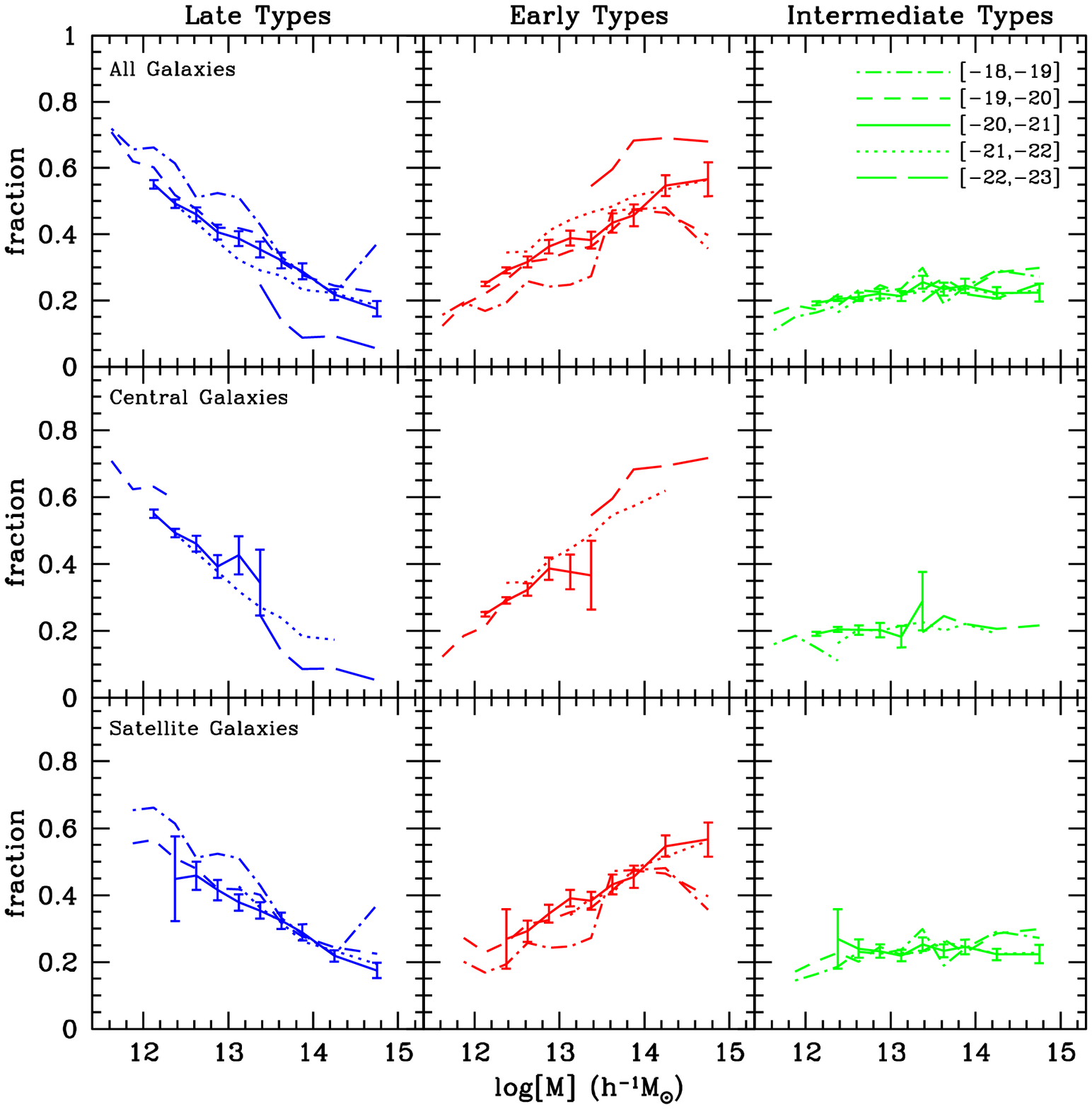,width=\hdsize}}
\caption{Same as Fig.~\ref{fig:typelum}, except that this time we plot
  the  type fractions  as function  of halo  mass for  five luminosity
  bins. The  values is square  brackets in the upper  right-hand panel
  indicate the  range of $^{0.1}M_r -  5 {\rm log} h$  used.  Note the
  strong, and smooth  halo mass dependence of the  early and late type
  fractions.   In   particular,  there   is  no  indication   for  any
  characteristic mass  scale.  Except  for the faintest  and brightest
  luminosity bins,  the fractions of  early and late type  galaxies at
  fixed  halo mass  are surprisingly  independent of  luminosity. Note
  that there is a weak indication that the mass dependence for central
  galaxies   is  stronger   than  for   satellite  galaxies.    As  in
  Fig.~\ref{fig:typelum}, the intermediate type fraction is completely
  independent of luminosity and halo mass, and is the same for central
  and satellite galaxies.  See text for a detailed discussion. }
\label{fig:typemass}
\end{figure*}

\subsection{Dependence on Halo Mass}
\label{sec:massdep}

We now investigate how galaxy type  depends on halo mass.  We start by
splitting the group sample in  six logarithmic mass bins and determine
how their  type fractions depend on  luminosity. For each  bin in mass
and luminosity the  late type fraction is defined  as the total number
of late  type galaxies  in that  bin, divided by  the total  number of
galaxies in that  bin (i.e., we do not average  the late type fraction
over  individual   haloes).   The  same  applies  to   the  early  and
intermediate types.

The results are  shown in the upper panels  of Fig.~\ref{fig:typelum}. 
In each mass bin the  late (early) type fraction decreases (increases)
with  increasing luminosity,  similar as  for the  entire sample  (cf. 
Fig.~\ref{fig:lumdep}).  Note,  however, that  in the range  $-19 \gta
^{0.1}M_r - 5  {\rm log} h \gta -21.5$  (indicated by vertical, dotted
lines), the luminosity  dependence is remarkably weak, for  all 6 mass
bins. For comparison, an $L^{*}$ galaxy has $^{0.1}M_r - 5 {\rm log} h
=  -20.44$  (Blanton  \etal  2003a),  so  that  this  magnitude  range
corresponds roughly  to $0.25 \lta  L/L^{*} \lta 2.5$.  In  Yang \etal
(2005c) we  found a similar result  from an analysis of  the early and
late type  fractions in 2dFGRS groups, despite  a different definition
of early and late types and the use of luminosities in the $b_J$-band,
rather than the $r$-band.

At fixed luminosity, the late and early type fractions depend strongly
on  halo mass: the  late type  fraction decreases  and the  early type
fraction  increases with  increasing halo  mass. Over  the  mass range
$10^{12} h^{-1} \Msun \lta M \lta 10^{15} h^{-1} \Msun$ both fractions
change by 30 to 40 percent, at all luminosities.  This is a reflection
of the  well known  morphology-density relation (e.g.,  Dressler 1980;
Postman  \&  Geller 1984;  Whitmore  1995;  Dom\'{i}nguez, Muriel  \&
Lambas 2001; Goto \etal 2003; Tanaka \etal 2004), but now expressed in
terms of halo mass rather than galaxy number density.

Panels in  the middle and lower  row show the  same results separately
for  central and  satellite  galaxies. As  expected, central  galaxies
mainly  occupy   the  bright  end   of  the  distribution.    In  the,
unfortunately  small,  magnitude range  where  satellites and  central
galaxies overlap, there  is a weak indication that  the early and late
type  fractions  of  central   galaxies  increase  and  decrease  with
luminosity, respectively,  while those  of the satellite  galaxies are
consistent   with  no  luminosity   dependence.  However,   given  the
(Poissonian) errors  we can  not rule out  that central  and satellite
galaxies  follow the  same trend;  a larger  data set  is  required to
investigate this in more detail.

\begin{figure*}
\centerline{\psfig{figure=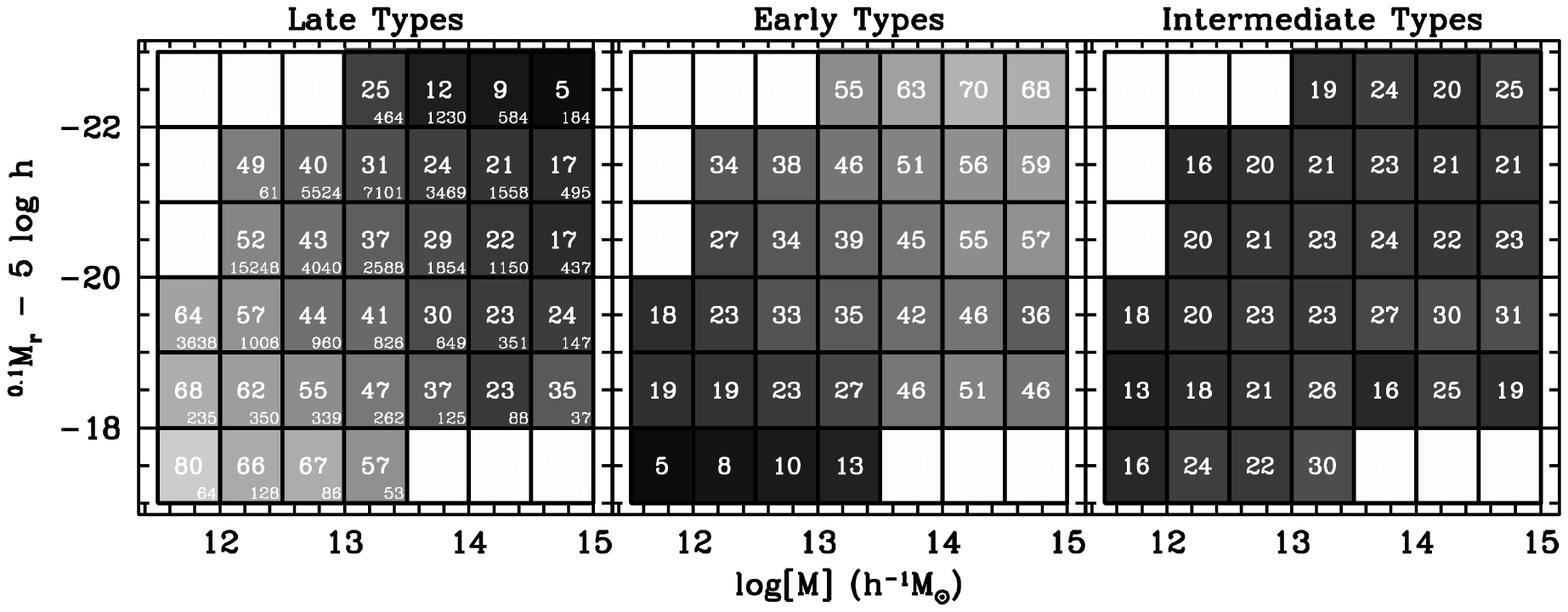,width=\hdsize}}
\caption{Galaxy type as a function of halo mass and luminosity for the
  galaxies in  our group catalogue. The  number in the  center of each
  cell indicates  the percentage of  late type galaxies  (left panel),
  early type  galaxies (middle panel),  and intermediate-type galaxies
  (right  panel).   Each  cell   is  colour-coded  according  to  this
  percentage, running from  black (0 percent) to white  (100 percent). 
  The number in  the lower-right corner of each  cell in the left-hand
  panel indicates  the total number  of galaxies in  the corresponding
  mass-luminosity bin.  Only cells with more than 50 galaxies in total
  are shown.}
\label{fig:res2D}
\end{figure*}

The  right-hand   panels  of  Fig.~\ref{fig:typelum}   show  that  the
intermediate  type fractions  are  once again  remarkably constant  at
$\sim  20$  percent; there  is  no  significant  dependence on  either
luminosity or halo mass, nor does it depend on whether the galaxy is a
central galaxy or a satellite galaxy. The implications of this for the
nature    of   intermediate   type    galaxies   are    discussed   in
Section~\ref{sec:inter}.
 
Fig.~\ref{fig:typemass} shows  these results  in a complementary  way. 
It  shows  the  type fractions  as  function  of  halo mass  for  five
different  magnitude bins.   For each  magnitude bin  we  only include
groups that  fall entirely  within the volume  limit, i.e.,  for which
{\it all}  members have $0.01 \leq  z \leq z_{\rm  max}$.  Whereas the
intermediate type  fraction, once again, shows no  significant mass or
luminosity dependence, the early  and late type fractions are strongly
mass dependent.  Most  importantly, we find the mass  dependence to be
remarkably smooth,  with no indication  at all for  any characteristic
mass  scale\footnote{The  only   apparent  exception  occurs  for  the
  brightest sample with  $-22 \geq {^{0.1}M}_r - 5  {\rm log}h > -23$,
  where the  late and early type  fractions seem to reveal  a break at
  around  $10^{14} h^{-1}  \Msun$.  However,  an investigation  of the
  Poissonian  errorbars (not shown)  suggests that  this break  is not
  significant.}

At fixed  halo mass, the  luminosity dependence is  surprisingly weak,
especially over  the magnitude  range $-19 \gta  {^{0.1}M}_r -  5 {\rm
  log}h \gta -22$.  The early and late type fractions only reveal some
luminosity dependence at the very bright and the very faint end of the
distribution (cf.  Fig.~\ref{fig:typelum}).

Panels in the middle and lower row of Fig.~\ref{fig:typemass} show the
various  type fractions  as  function  of halo  mass  for central  and
satellite galaxies, respectively.  There is  a weak hint that the mass
dependence  is   stronger  for  central  galaxies   (just  like  their
luminosity  dependence  is  stronger, see  Fig.~\ref{fig:lumdep}).   A
confirmation of this  trend, however, has to await  a larger sample of
(SDSS) data.

Note  that  the  functional  form  of the  mass  dependence  at  fixed
luminosity  is  very  similar  for  all  magnitude  bins  considered.  
Similarly, the  functional form of the luminosity  dependence at fixed
halo mass is very similar for  all mass bins.  This suggests a simple,
separable form  for the early and  late type fractions  as function of
luminosity  and mass,  i.e.,  $f_{\rm late}(L,M)  =  g(L) h(M)$,  with
$g(x)$ and $h(x)$ two (monotonic) functions. Such a separable form was
adopted  by van  den Bosch  \etal (2003)  and Cooray  (2005)  in their
studies of the conditional luminosity functions of early and late type
galaxies in the 2dFGRS. The results presented here provide support for
these functional forms, albeit in retrospect.

Finally, for completeness, Fig.~\ref{fig:res2D} shows the same results
once more, but now  in a two-dimensional representation. The grayscale
represents the fraction of late, early, and intermediate type galaxies
in  each   mass-luminosity  bin.   The  reader  can   read  off  these
percentages (big, white number in the center of each cell), as well as
the total number of galaxies in each bin (small, white number in lower
right corner of each cell).

\begin{figure}
\centerline{\psfig{figure=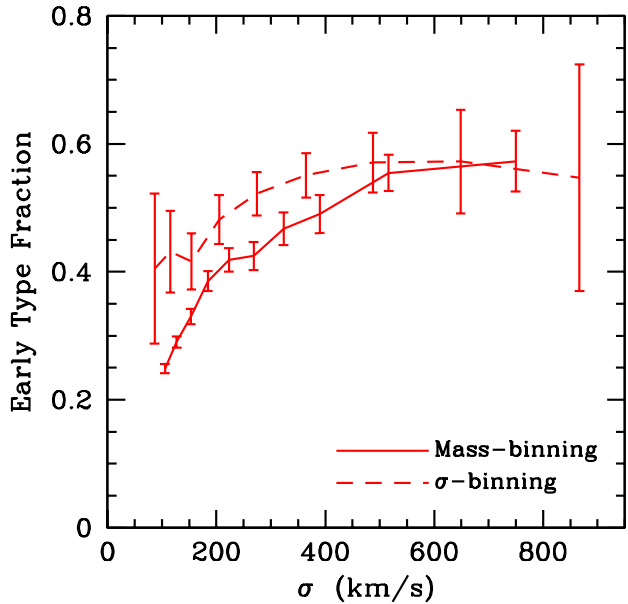,width=\hssize}}
\caption{The early-type fraction as function of group velocity
  dispersion for galaxies with $-20  \geq {^{0.1}M}_r - 5 {\rm log}h >
  -22$. Solid lines use our mass estimator (based on group luminosity)
  converted to velocity dispersion  using equation (A5).  Dashed lines
  use a binning based on  the actual velocity dispersion of the member
  galaxies.  Note that the latter predicts a significantly weaker mass
  dependence than the former.}
\label{fig:comp}
\end{figure}

Our finding that the late type fraction decreases with increasing halo
mass  is in agreement  with previous  results from  Mart\'{i}nez \etal
(2002) and Yang \etal (2005c). On the other hand, Tanaka \etal (2004),
de  Propris  \etal (2004)  and  Balogh  \etal  (2004b) find  {\it  no}
significant  dependence of  the late  or  early type  fraction on  the
velocity  dispersion of massive  groups and  clusters.  There  are two
reasons for  this apparent  discrepancy. First of  all, our  sample is
significantly  larger than that  of previous  studies. This  not only
results  in significantly  smaller errorbars,  but also  allows  us to
consider a much larger dynamic  range in halo masses.  Secondly, as we
show in Appendix~B, using the  velocity dispersion as a mass estimator
naturally  tends to  smear  out  the mass  dependence.   This is  also
illustrated  in  Fig.~\ref{fig:comp},  were  we  plot  the  early-type
fraction of galaxies with $-20 \geq  {^{0.1}M}_r - 5 {\rm log}h > -22$
(using  a volume  limited sample)  as function  of the  group velocity
dispersion.  The  solid lines use  our mass estimator (based  on group
luminosity),  converted to  velocity dispersion  using equation  (A5). 
Dashed lines use a binning  based on the actual velocity dispersion of
the member galaxies. Only groups  with 6 members or more are included,
although the results look similar when using all groups with 3 members
or more.   Note that  over the  range $350 \kms  \lta \sigma  \lta 850
\kms$, which is the range used in Balogh \etal (2004a), the early-type
fraction is  basically flat when  using the velocity dispersion  of the
member  galaxies. This  explains the  discrepancy between  the results
presented here and those in the previous studies listed above.

Finally,  there have been  a number  of recent  studies that  used the
clustering properties of early and late type galaxies to constrain the
type  fractions as function  of halo  mass. Magliocchetti  \& Porciani
(2003), van den Bosch \etal  (2003), and Collister \& Lahav (2005) all
used  the  two-point correlation  functions  of  early  and late  type
galaxies in  the 2dFGRS to  infer that the  late type fraction  has to
decrease smoothly with halo  mass, in good, qualitative agreement with
the  results presented  here. See  also Cooray  (2005) for  a somewhat
different analysis,  but with the  same result.  An exception  to this
behavior was  found by Zehavi \etal  (2004), who inferred  a late type
fraction from  the correlation functions  extracted from the  SDSS DR2
that decreases with halo mass  down to a minimum at $10^{13.5} \Msun$,
followed by  a subsequent increase. Unfortunately,  as demonstrated in
van den Bosch \etal (2003), the uncertainties on the type fractions as
inferred solely from the clustering  data are fairly large, so that we
do not  consider the results of  Zehavi \etal (2004) to  be in serious
conflict with those presented here.

\subsection{Dependence on Halo-centric Radius}
\label{sec:raddep}

\begin{figure*}
\centerline{\psfig{figure=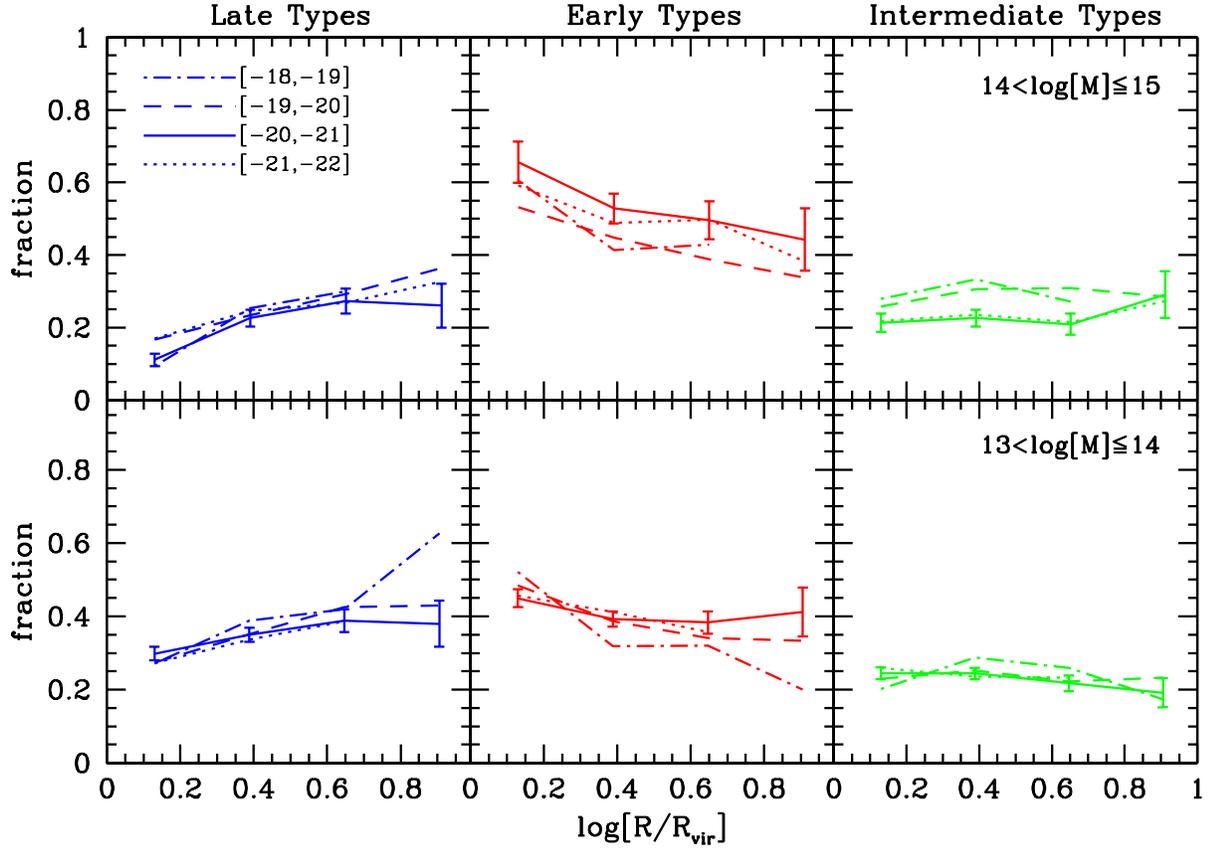,width=0.9\hdsize}}
\caption{The fractions of satellite galaxies that are late type
  (left-hand  panels), early  type (middle  panels),  and intermediate
  type  (right-hand panels) as  function of  the projected  radius $R$
  from  the (luminosity-weighted)  group center  (in units  of $R_{\rm
    vir}$).  Results  are shown for  haloes in two mass  ranges (upper
  and  lower  panels),  and   for  four  bins  in  absolute  magnitude
  (different line styles, as indicated  in the upper left-hand panel). 
  We only  show results for bins  in radius, magnitude,  and mass that
  contain at  least 50 galaxies in  total.  For clarity,  we only show
  (Poissonian)  errorbars for one  of the  four magnitude  bins.  Note
  that the fraction of  late (early) type satellite galaxies increases
  (decreases)  significantly with radius,  independent of  luminosity. 
  However,  the  fraction of  intermediate  type  satellites does  not
  reveal any radius, luminosity, or mass dependence.}
\label{fig:raddep}
\end{figure*}

Thus far  we have  focussed on the  luminosity and mass  dependence of
galaxy type fractions.  Here we address the dependence on halo-centric
radius, i.e.,  we explore  the environment dependence  on scales  $R <
R_{\rm vir}$.   In order to  be able to discriminate  type segregation
from luminosity  segregation we investigate the  radial dependence for
four magnitude bins.  As above,  for each magnitude bin we construct a
volume  limited sample,  in which  we  only include  haloes that  fall
entirely within this volume.  For each galaxy we compute the projected
distance,  $R$, to  the  (luminosity weighted)  group  center, at  the
(luminosity weighted) redshift of the group.  In order to allow groups
of different masses  to be stacked together, we  normalize these radii
to the  group's virial radius $R_{\rm  vir}$\footnote{Virial radii are
  computed from  our group masses,  which we convert to  virial masses
  using the  relation between halo  mass and concentration  in Bullock
  \etal  (2001).}   Results  are  shown in  Fig.~\ref{fig:raddep}  for
groups  in  two separate  mass  ranges.   Since  central galaxies  are
special,   we  have  excluded   them  from   our  analysis,   so  that
Fig.~\ref{fig:raddep} only reflects the type fractions of satellites.

In  agreement with  previous studies  (e.g., Postman  \&  Geller 1984;
Biviano  \etal 2002;  Dom\'{i}nguez  \etal 2002;  Girardi \etal  2003;
G\'{o}mez \etal 2003;  Goto \etal 2003; Goto \etal  2004) we find that
the late type fractions increase  towards the outskirts of the groups. 
Since this trend  is virtually identical for all  four magnitude bins,
it is  not a  reflection of luminosity  segregation (see  also Girardi
\etal 2003).

\begin{figure*}
  \centerline{\psfig{figure=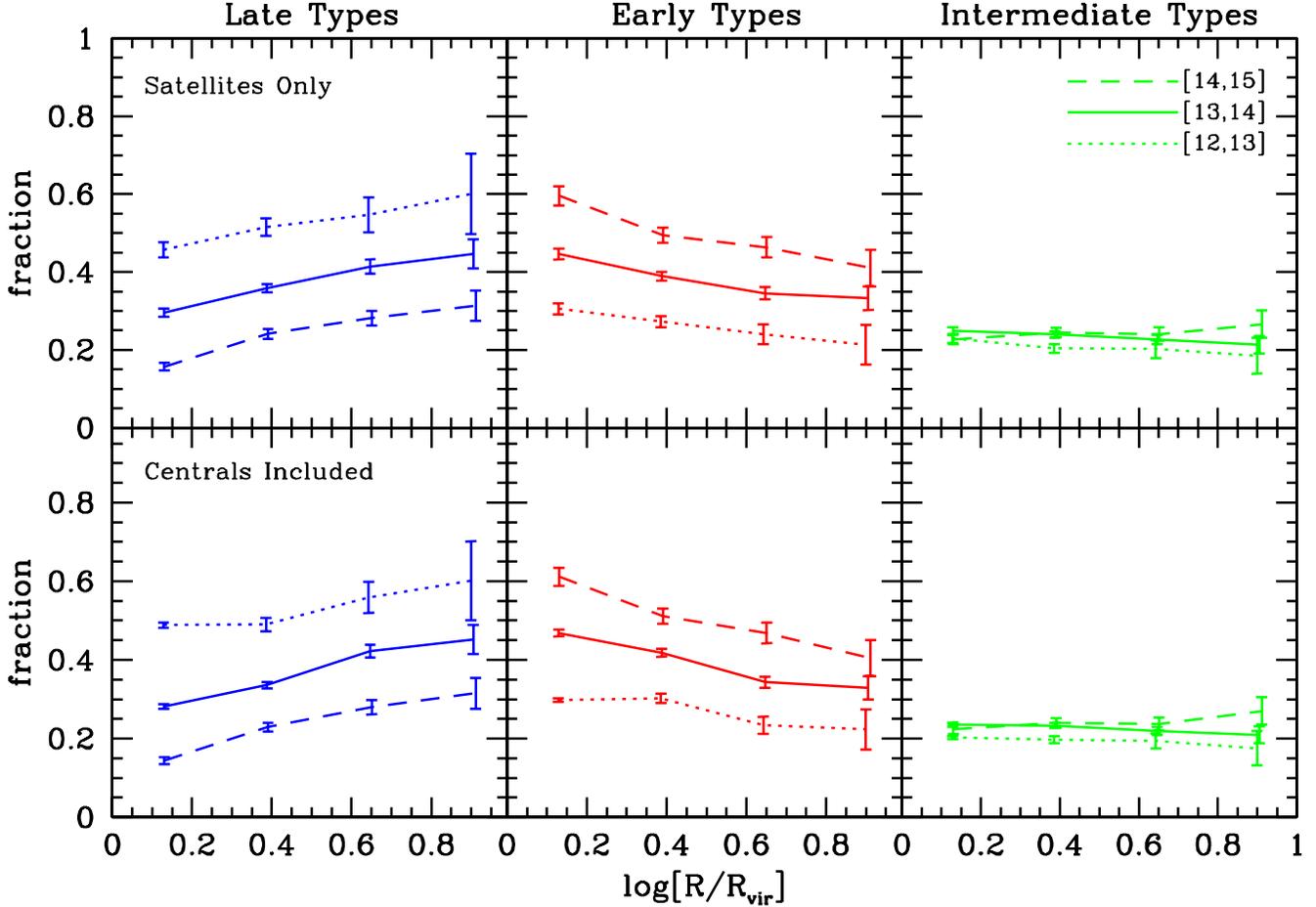,width=\hdsize}}
\caption{The upper panels are the same as Fig.~\ref{fig:raddep} except 
  that this  time we consider  all galaxies in the  magnitude interval
  $-23 \leq  ^{0.1}M_r - 5  {\rm log}h \leq  -18$ and in  the redshift
  range  $0.01 \leq z  \leq 0.2$.   Results are  shown for  three mass
  bins.  The values  in square brackets in the  upper right-hand panel
  indicate  the values  of  ${\rm log}(M)$  (in  $h^{-1}\Msun$) used.  
  Except for an offset, which  reflects the halo mass scaling shown in
  Fig.~\ref{fig:typemass},  the radial  dependence  is independent  of
  halo mass.  For comparison,  the lower panels  reveal the  same type
  fractions,  but  this  time  central  galaxies  are  included.  This
  introduces a weak mass dependence, in that lower mass haloes seem to
  reveal  a weaker  dependence  on  radius. See  text  for a  detailed
  discussion.}
\label{fig:radmass}
\end{figure*}

Having established that there  is no significant luminosity dependence
at fixed halo mass and radius,  we now increase the signal to noise by
computing the type fractions over the entire magnitude range from $-23
\leq ^{0.1}M_r - 5 {\rm log}  h \leq -18$ and over the entire redshift
range $0.01 \leq z \leq 0.20$.  Note that this is not a volume limited
sample.  However,  since we  have shown that  there is  no significant
luminosity dependence, Malmquist bias should not affect these results.
We  have  verified  that  using  a $1/V_{\rm  max}$  weighting  yields
virtually identical  results.  Due  to the increase  in the  signal to
noise  we can  now also  probe the  radial dependence  in  haloes with
masses  in the range  $10^{12} h^{-1}  \Msun <  M \leq  10^{13} h^{-1}
\Msun$   (results    for   this   mass   bin   are    not   shown   in
Fig.~\ref{fig:raddep} because they are  too noisy).  Results are shown
in  the   upper  panels  of  Fig.~\ref{fig:radmass}.    Except  for  a
normalization offset,  which reflects the halo mass  dependence of the
type fractions, {\it  the radial dependence is the  same for all three
  mass bins}.  In all cases, the late type fraction increases by $\sim
15$ percent  going from $R  \simeq 0.1 R_{\rm  vir}$ to $R  \simeq 0.9
R_{\rm vir}$.  Although this may  seem a relatively small increase, it
is  important to  realize that  we  observe the  radial dependence  in
projection.  Furthermore,  typical orbits  in dark matter  haloes have
fairly  large apo-  to pericenter  ratios  of $6:1$  or larger  (e.g.,
Ghigna \etal 1998; van den  Bosch \etal 1999), which together with the
projection makes the  observed trend appear much weaker  than the real
trend.

Our result  that the radial  trend is independent  of halo mass  is in
conflict  with Dom\'{i}nguez  \etal (2002)  who, using  the  100K data
release of the  2dFGRS, found no significant radial  dependence of the
late type  fraction in  haloes with $M  \lta 10^{13.5} h^{-1}  \Msun$. 
There are two  reasons for this discrepancy. First  of all, our sample
is significantly larger, resulting  in smaller errorbars. Secondly, as
far as  we can tell,  Dom\'{i}nguez \etal (2002) included  the central
galaxies in their analysis.  If we  do the same, we obtain the results
shown  in the lower  panels of  Fig.~\ref{fig:raddep}.  Note  that the
inclusion of  central galaxies slightly boosts the  late type fraction
in the  inner most  radial bin, especially  for low mass  haloes. This
reduces the overall  radial trend, and for the  mass bin with $10^{12}
h^{-1} < M \leq 10^{13} h^{-1}  \Msun$ the data is now consistent with
no  significant  radial dependence,  in  agreement with  Dom\'{i}nguez
\etal (2002).  Since central galaxies are special in many respects, we
feel,  however,  that it  is  more  appropriate  to study  any  radial
dependence using satellite galaxies only.

Finally we note that the intermediate type fraction is, in addition to
being  independent   of  galaxy   luminosity  and  group   mass,  also
independent  of halo-centric  radius (in  all mass  bins, and  for all
luminosity bins).  Thus, a  randomly selected galaxy, whether faint or
bright, whether in a low mass  halo or a cluster, and whether close or
far  from the group/halo  center, has  a $\sim  20$ percent  chance of
being  an intermediate  type galaxy  (see  Section~\ref{sec:inter} for
discussion).

\subsection{Dependence on Central Galaxy Type}
\label{sec:typedep}

\begin{figure*}
\centerline{\psfig{figure=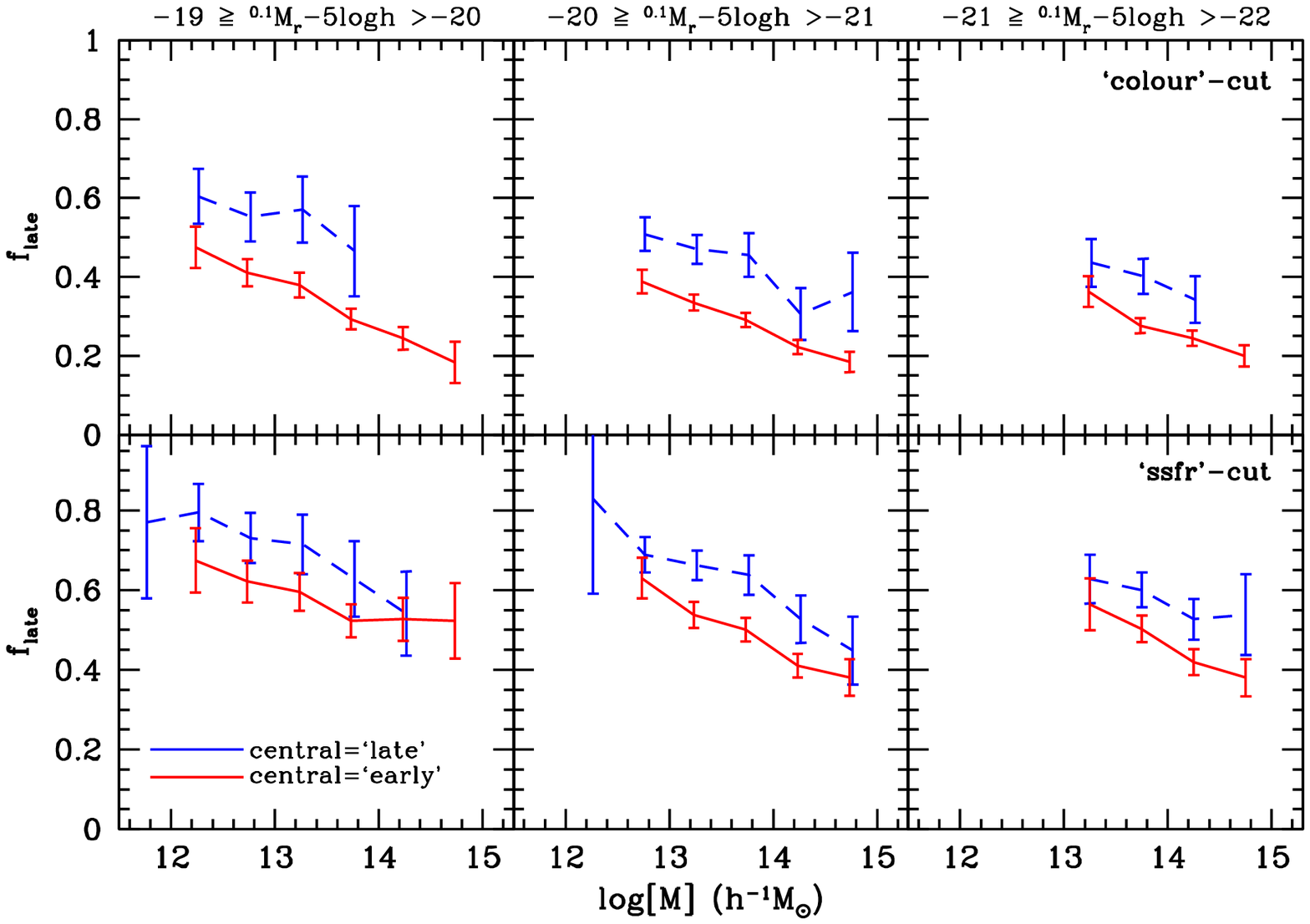,width=0.9\hdsize}}
\caption{The late type fraction of satellites galaxies as function of
  halo mass  for haloes with a  central early type  galaxy (red, solid
  curves)  and a  central  late  type galaxy  (blue,  dashed curves).  
  Results  are shown for  three different  volume limited  samples, as
  indicated.  In the  upper panels, galaxy type is  determined using a
  colour cut (equation~[\ref{colmagcrit}]),  while in the lower panels
  a  cut based  on  the SSFR  (equation~[\ref{ssfrmagcrit}]) has  been
  used. Results are only shown  for mass-luminosity bins with at least
  50 galaxies  in total, and errorbars denote  Poissonian errors. Note
  that haloes  with a  late type central  galaxy have  a significantly
  higher fraction  of late type  satellites than haloes with  an early
  type central galaxy, a phenomenon we term `galactic conformity'.}
\label{fig:seg}
\end{figure*}

Next  we  investigate whether  the  properties  of satellite  galaxies
correlate with those of their central galaxy. Fig.~\ref{fig:seg} plots
the late  type fraction as function  of halo mass  for three different
magnitude bins (again computed  using volume limited samples). Here we
use a different type classification than in the rest of this paper. In
the upper panels we split galaxies in late and early types only (i.e.,
no intermediate types are defined here), using the colour cut given by
equation~(\ref{colmagcrit}). Galaxies that are bluer than this cut are
termed late  types. In the  lower panels the  split in early  and late
types  is  based on  the  SSFR  cut  of equation~(\ref{ssfrmagcrit}).  
Galaxies with  a SSFR  that is  higher than this  cut are  called late
types.   In  each  panel  in  Fig.~\ref{fig:seg}  blue,  dashed  lines
indicate the late  type fraction of satellites in  haloes in which the
central galaxy  is also a late  type.  Red, solid  lines correspond to
haloes with an early type central galaxy.  Note that the luminosity of
the central galaxy is not  restricted to fall within the magnitude bin
indicated.

As is evident from Fig.~\ref{fig:seg}, haloes with a late type central
galaxy have  a significantly higher  fraction of late  type satellites
than haloes  of the same mass but  with an early type  central galaxy. 
This  difference is  evident  over  the entire  ranges  of masses  and
luminosities explored. Apparently, satellite galaxies `know' about the
properties  of their central  galaxy. 

This phenomenon, which we term  `galactic conformity', is a new result
that has  not been noticed  before. Some studies, however,  have found
correlations that point in  the same direction. Wirth (1983), studying
the galaxy  content of groups and clusters  using photographic plates,
noted  that the direct  environment of  elliptical galaxies  contain a
higher fractions of early types than the average of the field. Hickson
\etal (1984),  studying compact groups, noticed that  if the brightest
galaxy is a spiral the fainter  group members also tend to be spirals. 
Ramella  \etal (1987),  analyzing the  morphological content  of loose
groups in the  catalogue of Geller \& Huchra  (1983), noticed that the
fraction  of  elliptical  galaxies  is  significantly  higher  if  the
first-ranked  group  member is  also  an  elliptical.   None of  these
studies, though, performed  the analysis as a function  of group mass. 
Since  the early  type  fraction  increases with  halo  mass for  both
central  and  satellite  galaxies (see  Section~\ref{sec:massdep}),  a
type-correlation between the central  galaxy and its satellites arises
naturally when using  a sample of groups that span a  range in masses. 
Indeed, Osmond \& Ponmon (2004), studying a sample of 60 galaxy groups
with existing  X-ray data, also  noticed that the spiral  fraction was
significantly higher  if the  brightest group galaxy  also was  a late
type.  The corresponding groups, however,  where found to have a lower
velocity dispersion  and no  detected X-ray emission,  suggesting that
they had a lower mass on  average. What is special about the `galactic
conformity' presented here, is that  such a correlation exists {\it at
  a  fixed  halo  mass},  and  for  satellites of  a  fixed  range  in
magnitudes.  This  finding puts  intriguing new constraints  on galaxy
formation     models,     which      we     briefly     address     in
Section~\ref{sec:conform}.

\subsection{The Correlation between Galaxy Properties and Halo Mass}
\label{sec:propmass}

\begin{figure*}
\centerline{\psfig{figure=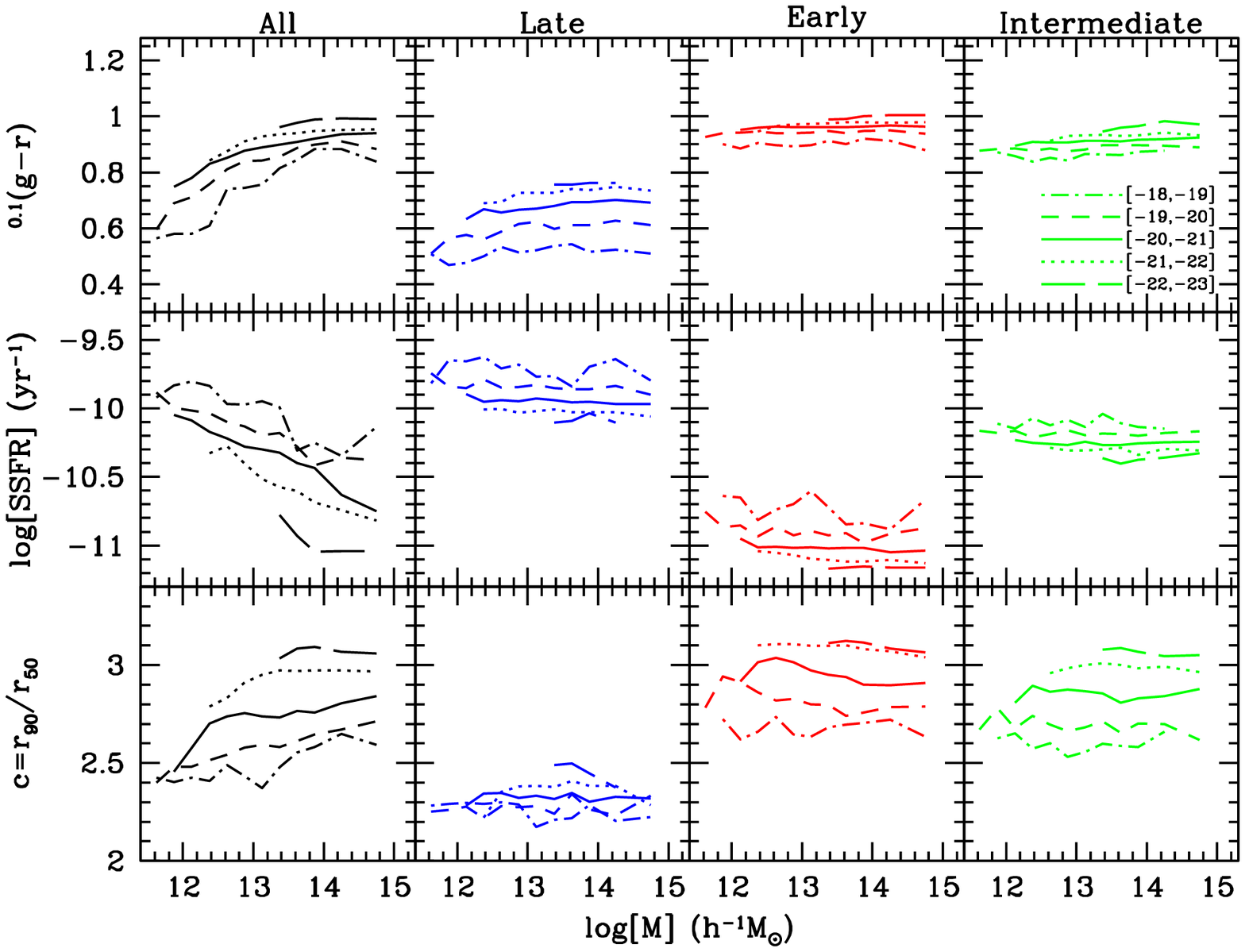,width=\hdsize}}
\caption{The {\it median} colour, SSFR, and concentration of galaxies 
  as function of halo mass.  Results are shown for five magnitude bins
  (as indicated), and separately for all galaxies, late type galaxies,
  early type galaxies and intermediate type galaxies. Results are only
  shown for mass-luminosity bins that  contain at least 20 galaxies in
  total.}
\label{fig:medprop}
\end{figure*}

Thus far we  have only focussed on the {\it  fractions} of early, late
and intermediate  type galaxies. We  now examine how the  {\it median}
colour, SSFR, and  concentration of galaxies scale with  halo mass. As
before, we  discriminate between  luminosity dependence and  halo mass
dependence by splitting the galaxy population in a number of magnitude
bins.  For  each bin  we construct a  volume limited sample,  and only
consider  groups that fall  entirely within  this volume.  Results are
shown in Fig.~\ref{fig:medprop}, which  plots the median colour (upper
panels), SSFR (panels in middle row), and concentration (lower panels)
as  function of  halo  mass\footnote{We have  also  examined the  {\it
    average}  properties (not shown),  and found  the relations  to be
  extremely similar.}.  Results are  shown for five magnitude bins and
separately for  all galaxies, late type galaxies,  early type galaxies
and intermediate type galaxies.

If we  first focus on the  relations for all galaxies  (panels in left
column),  one  notices  that  the  correlations of  all  three  galaxy
properties with halo  mass are fairly weak {\it  at fixed luminosity}. 
To make this a bit more quantitative, we estimate the gradients of the
median  properties as  function of  mass at  fixed luminosity,  and as
function of  luminosity at  fixed mass.  For  the luminosity  and mass
dependence of the median colour we find
\begin{equation}
\label{coldep}
{\rmd ^{0.1}(g-r) \over \rmd \log M}\Big{|}_L \approx +0.06 
\;\;\;\;\;\;\;\;\;\;\,
{\rmd ^{0.1}(g-r) \over \rmd \log L}\Big{|}_M \approx +0.09
\end{equation}
For the SSFR these gradients are
\begin{equation}
\label{sfrdep}
{\rmd \log {\rm SSFR} \over \rmd \log M}\Big{|}_L \approx -0.20 
\;\;\;\;\;\;\;\;
{\rmd \log {\rm SSFR} \over \rmd \log L}\Big{|}_M \approx -0.35
\end{equation}
and for the concentration we find
\begin{equation}
\label{condep}
{\rmd c \over \rmd \log M}\Big{|}_L \approx +0.07 
\;\;\;\;\;\;\;\;\;\;\;\;\;\;\,
{\rmd c \over \rmd \log L}\Big{|}_M \approx +0.25
\end{equation}
Although these numbers are fairly rough estimates, it is clear that in
all three  cases the luminosity  dependence is stronger than  the halo
mass dependence (when both luminosity  and mass are expressed in solar
units).

Note  that this  contrasts strongly  with the  type  fractions, which
depend   more  strongly  on   halo  mass   than  on   luminosity  (see
Section~\ref{sec:massdep}).   We can  reconcile this  with  the strong
luminosity dependence of the median  colour and SSFR by realizing that
the cuts in colour and SSFR used to define the galaxy types scale with
luminosity according to
\begin{equation}
\label{bimodal}
{\rmd ^{0.1}(g-r) \over \rmd \log L} = +0.08 
\;\;\;\;\;\;\;\;\;\;\;
{\rmd \log {\rm SSFR} \over \rmd \log L} = -0.24
\end{equation}
(cf. equations~[\ref{colmagcrit}] and~[\ref{ssfrmagcrit}]).  Note that
these gradients  are comparable to  those of the median  properties at
fixed mass. This shows that at  fixed halo mass, the median colour and
SSFR  scale   roughly  with  luminosity   in  the  same  way   as  the
corresponding bimodality  scales. The fractions of  galaxies on either
site  of  this  bimodality  scale,  however,  only  depend  weakly  on
luminosity at fixed halo mass.

As    is   evident    from   the    three   right-hand    columns   of
Fig.~\ref{fig:medprop}  the median  properties  of a  galaxy of  given
luminosity {\it and type} are virtually independent of halo mass.  The
mass  dependence  of the  median  properties  of  {\it all}  galaxies,
therefore, owes entirely to the mass dependence of the type fractions.
In  as far as  halo mass  is a  reliable proxy  for the  local surface
density of  galaxies, this is  in agreement with Balogh  \etal (2004a)
and Tanaka \etal (2004) who  found that although the {\it fraction} of
star forming  galaxies (defined according  to the equivalent  width of
the  H$\alpha$  line)  depends  strongly  on  $\Sigma_5$,  the  median
equivalent width  of star forming galaxies (those  with ${\rm EW}({\rm
  H}\alpha) >  4${\AA}) does not show  any $\Sigma_5$-dependence.  Our
results also  agree with  those of Kauffmann  \etal (2004),  who found
that the concentration parameter  of galaxies is independent of galaxy
number density at fixed stellar mass.

\subsection{Conditional Probability Distributions}
\label{sec:condmass}

\begin{figure*}
\centerline{\psfig{figure=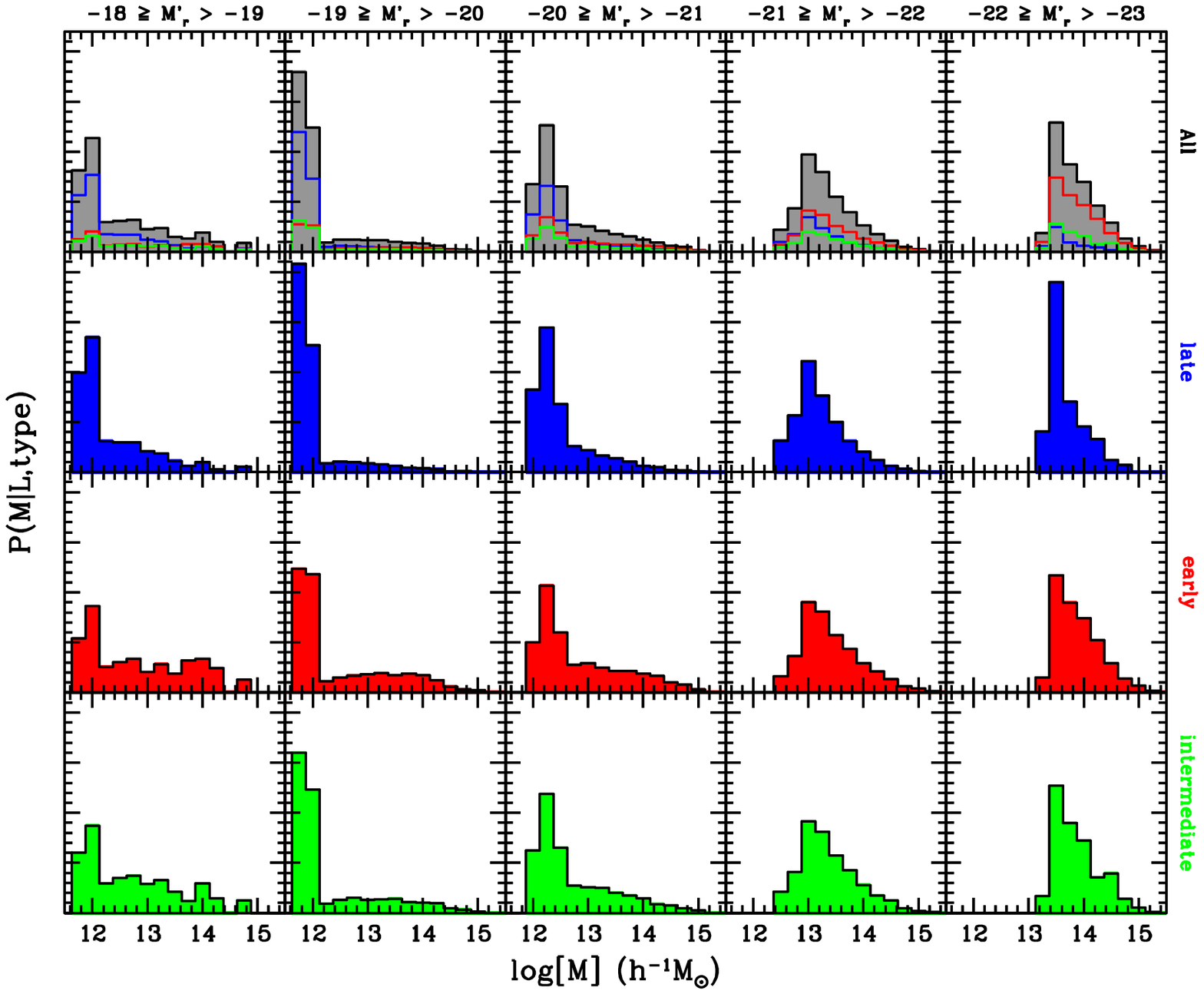,width=\hdsize}}
\caption{The conditional probability distribution $P(M \vert
  L,{\rm  type})$  that a  galaxy  of  given  luminosity $L$  (in  the
  $^{0.1}r$-band)  and given  type  resides in  a  halo of  mass $M$.  
  Results are shown for five  magnitude bins (indicated at top of each
  column, with $M'_r = ^{0.1}M_r -  5 {\rm log} h$) and for late types
  (panels  in second  row),  early  types (panels  in  third row)  and
  intermediate types (panels  in bottom row). The upper  row of panels
  plots the conditional probability  distribution $P(M \vert L)$ (gray
  scale).  The blue, red and green histograms in these panels indicate
  the  contributions  to  $P(M  \vert  L)$  due  to  late,  early  and
  intermediate types,  respectively. The  total number of  galaxies in
  each distribution,  $N$, is  indicated in the  upper left  corner of
  each panel.}
\label{fig:pltype}
\end{figure*}

The type fractions  and medians discussed thus far  are simple scalars
expressing   some    properties   of   the    underlying   probability
distributions. For completeness, we now present, for some illustrative
cases, these full distributions. First we split our sample of galaxies
(those  that have  been  assigned  to groups)  according  to type  and
luminosity  (using  five volume  limited  magnitude  bins).  For  each
galaxy in each luminosity-type bin we look up the mass of the group of
which  it  is a  member.   Fig.~\ref{fig:pltype}  plots the  resulting
conditional  mass functions $P(M  \vert L,{\rm  type})$, with  $L$ the
luminosity in the $^{0.1}r$-band.   The histograms in the upper panels
show $P(M  \vert L)$.  As  expected, bright galaxies always  reside in
massive  haloes.  The  conditional mass  function for  faint galaxies,
however, reveals  a bimodal  distribution: a narrow  peak at  low halo
masses, corresponding  to central galaxies,  and a very broad  wing to
high halo masses, corresponding  to satellite galaxies.  Note that the
functional form  of $P(M \vert L)$  derived here is  in good agreement
with  predictions   based  on  the   conditional  luminosity  function
presented in Yang  \etal (2003) and Cooray (2005).   The blue, red and
green  histograms in the  upper panels  indicate the  contributions to
$P(M  \vert L)$  due to  late, early  and intermediate  type galaxies,
respectively.  In  agreement  with  the results  shown  above,  bright
galaxies in massive haloes  are predominantly early types, while faint
galaxies in low mass haloes  are dominated by late types. However, one
can also see that those faint galaxies that reside in the most massive
haloes are more likely to be an early type.

The  latter is  more evident  when one  compares the  conditional mass
functions of early and late type  galaxies, shown in the panels in the
middle two  rows. For  faint galaxies, $P(M  \vert L,{\rm  late})$ and
$P(M \vert L,{\rm  early})$ are clearly different, in  that the former
is clearly  more skewed towards low  $M$.  This implies  that a faint,
early type  galaxy lives in a  halo that, on average,  is more massive
than a halo  hosting a late type galaxy of  the same luminosity.  This
is in good agreement with  other studies. In particular, Blanton \etal
(2005b)  studied  the  relationship  between environment  and  various
properties of galaxies in the SDSS. They computed the {\it mean} local
overdensity  as   function  of  both  luminosity   and  several  other
parameters,  including  colour   and  Sersic  index.   Although  their
overdensities are  measured using  a fixed metric  scale of  $1 h^{-1}
\Mpc$,  which,  as  we  have  argued  in  Section~\ref{sec:intro},  is
difficult to interpret in terms  of halo masses, their results paint a
very similar picture: blue faint  galaxies live in low density regions
(i.e., are central galaxies in  their own, low mass haloes), while red
faint galaxies reside in regions with a similar overdensity as that of
red bright galaxies  (i.e., they are satellite galaxies  in clusters). 
This is also consistent  with clustering data.  In particular, Norberg
\etal (2002)  have shown  that the correlation  length of  faint early
types  is much  higher than  that of  late type  galaxies of  the same
luminosity, indicating that they live in more massive haloes.

\begin{figure*}
  \centerline{\psfig{figure=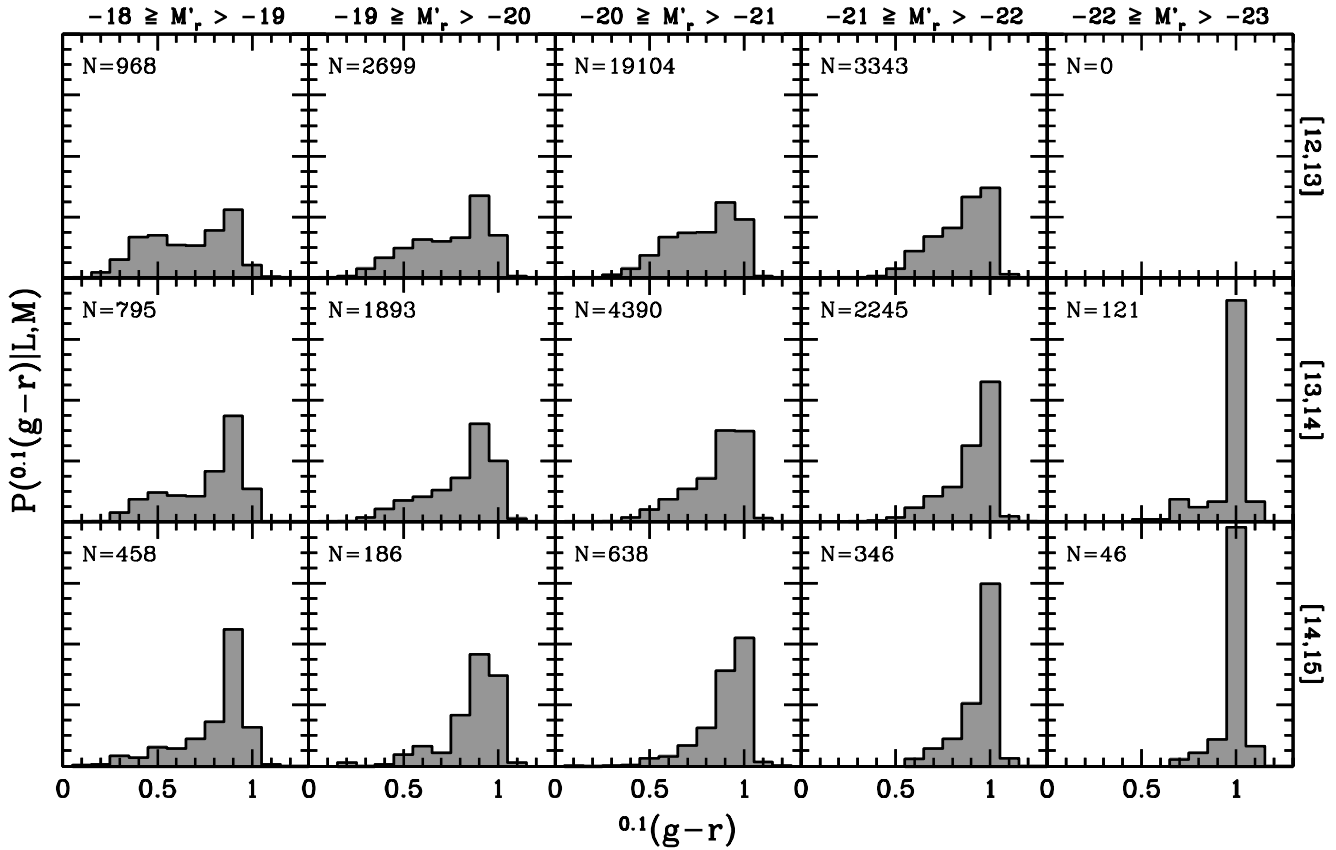,width=\hdsize}}
\caption{The conditional probability  distribution 
  $P(^{0.1}(g-r)  \vert L,M)$ for  three different  bins in  halo mass
  (values in  square brackets on the  right site of  each row indicate
  the range of  $\log[M]$ used) and five different  bins in luminosity
  (indicated at  top of each column,  with $M'_r = ^{0.1}M_r  - 5 {\rm
    log} h$). The total number  of galaxies in each distribution, $N$,
  is indicated in the upper left corner of each panel.}
\label{fig:colprob}
\end{figure*}

\begin{figure*}
  \centerline{\psfig{figure=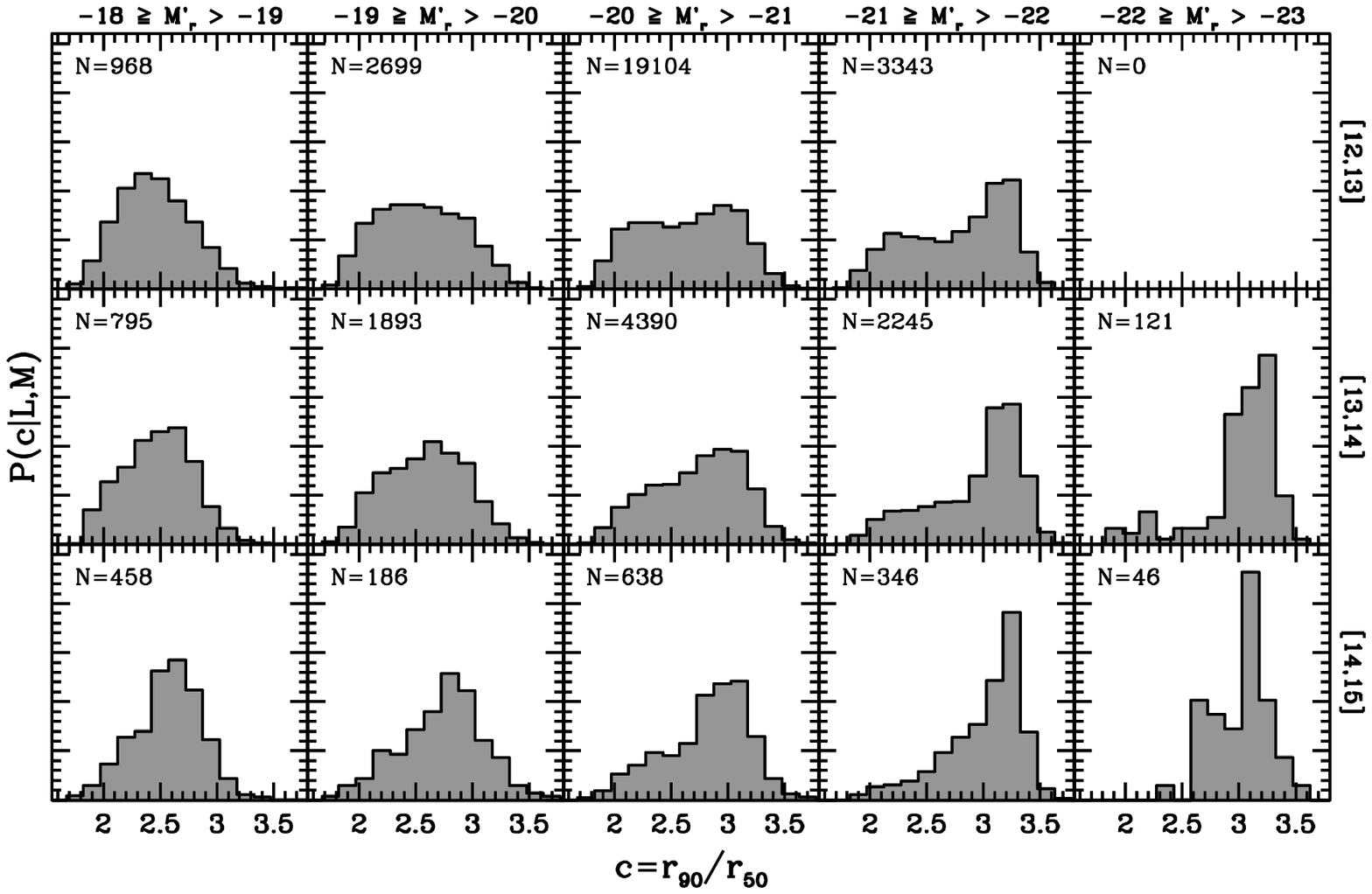,width=\hdsize}}
\caption{Same as Fig.~\ref{fig:colprob} except that this time we plot
  the  conditional  probability distribution  $P(c  \vert L,M)$,  with
  $c=r_{90}/r_{50}$ the galaxy concentration.}
\label{fig:conprob}
\end{figure*}

Fig.~\ref{fig:colprob}  plots  the  conditional  colour  distributions
$P(^{0.1}(g-r)  \vert L,M)$  for  three  bins in  halo  mass and  five
(volume limited) bins  in absolute magnitude.  There is  a clear trend
that the fraction  of red galaxies increases with  both luminosity and
with halo mass,  in agreement with the results  presented above.  Note
also that, at fixed halo  mass, the colour distributions for $-19 \geq
^{0.1}M_r - 5 {\rm log} h >  -20$ and for $-20 \geq ^{0.1}M_r - 5 {\rm
  log} h > -21$ are  remarkably similar, consistent with the fact that
the  galaxy type  fractions are  independent of  luminosity  over this
magnitude     range    (cf.      Fig.~\ref{fig:typelum}).     Finally,
Fig.~\ref{fig:conprob} plots the distributions of galaxy concentration
conditional on luminosity and  halo mass.  These nicely illustrate how
the  average   concentration  increases   with  both  halo   mass  and
luminosity, as  already shown in  Fig.~\ref{fig:medprop}. Although all
main trends visible in Figs.~\ref{fig:pltype} -- \ref{fig:conprob} are
already evident  from the previous discussion based  on type fractions
and  median  properties, the  full  distributions  shown here  contain
useful, additional  information not evident from the  fractions or the
means.

\section{Discussion} 
\label{sec:disc}

\subsection{Implications for Galaxy Formation \& Evolution}
\label{sec:galform}

In the current paradigm of galaxy formation, galaxies form in extended
dark matter haloes. In the pure `nature' scenario, the properties of a
galaxy  depend only  on the  mass and  formation history  of  the dark
matter halo  in which it  resides. However, a galaxy  also experiences
interactions of various kinds  with its environment. Examples of these
are  ram-pressure  stripping, strangulation,  and  galaxy harassment.  
These, and other, `nurture' processes  may also play an important role
in setting the final properties of a galaxy.

Ever since  the discovery that galaxy properties  correlate with their
environment,  their has  been an  ongoing  debate as  to the  relative
importance  of nature  versus nurture  processes in  regulating galaxy
properties. In  this paper  we have analyzed  how a variety  of galaxy
properties  depend on  halo  mass,  using a  sample  of $\sim  90,000$
galaxies distributed over $\sim  53,000$ haloes (galaxy groups). These
results provide a testbed for comparison with galaxy formation models,
and may provide important insights regarding the nature-versus-nurture
debate.

Unfortunately,  many poorly understood,  intertwined processes  play a
role in galaxy formation, so  that an interpretation of our results is
far from straightforward.  For  example, although the mere presence of
a correlation between galaxy properties and environment is often taken
as  evidence  for  a  dominant  role of  `nurture'  processes,  it  is
important to realize that many,  if not all, of these correlations can
equally well be explained within a pure `nature' scenario (see below).
This makes it extremely  difficult to discriminate between the various
physical processes.  Below we briefly discuss some of these processes,
and  emphasize how  their  (often crude  and speculative)  predictions
compare to the results presented above.

\subsubsection{The nature scenario}
\label{sec:nature}

In  the  `nature'  scenario,  the  global  properties  of  the  galaxy
population owe  mainly to the  formation history of their  dark matter
haloes.   During quiescent  growth  phases  gas can  cool  and form  a
centrifugally supported  disk. Star formation  slowly and continuously
converts the gas  into stars, resulting in a  typical late-type galaxy
with blue colors and an ongoing SFR. During a major merger of two dark
matter haloes, the (central) galaxies  are likely to merge as well due
to dynamical friction. If their  mass ratio is sufficiently small, the
outcome  of this  merger event  will most  likely resemble  a spheroid
(e.g., Toomre \& Toomre 1972), while most gas is likely to be consumed
in a  starburst (e.g., Mihos \&  Hernquist 1996). If  new accretion of
gas can  somehow be  prevented, for example  by invoking  AGN feedback
(Croton \etal 2005), the resulting galaxy will quickly become `red and
dead',  characteristic  of a  genuine  early-type.   If, however,  the
accretion of  new gas can  not be prevented,  a new disk may  start to
grow around  the spheroid, slowly  transforming the early type  into a
late type.

This is the standard  picture adopted in virtually all semi-analytical
models for galaxy formation.  When assigning galaxy types according to
their bulge-to-disk  ratio, these models yield  an increasing fraction
of  early  types  with   increasing  halo  mass  and  with  decreasing
halo-centric radius  (Diaferio \etal 2001; Okamoto  \& Nagashima 2001;
Springel \etal 2001; Berlind \etal 2003), all in qualitative agreement
with observations.   This suggests that  the global morphology-density
relation  is built  in at  a  very fundamental  level in  hierarchical
formation theories  and can be explained within  the `nature' scenario
(see  also Evrard,  Silk \&  Szalay 1990).   However, all  models have
problems in trying to match  the radial dependence of S0s, which seems
to require additional (`nurture') processes.

In addition, a more  detailed comparison between the model predictions
and the  results presented here  indicates another potential  problem. 
Since  galaxy-galaxy merging  is  inefficient in  massive haloes,  all
semi-analytical  models  predict  a  `saturation' of  the  early  type
fraction in  haloes above a certain  mass. For example,  the models of
Diaferio  \etal  (2001)  predict  that  the fraction  of  early  types
(defined according to bulge-to-disk ratio) increases with halo mass up
to $\sim 10^{13.5} h^{-1} \Msun$,  after which the early type fraction
reveals a modest decline. This is inconsistent with our results, which
show that the early type fraction continues to decrease up to the most
massive  haloes analyzed  ($M \simeq  10^{15} h^{-1}  \Msun$). Berlind
\etal   (2003)  and   Zheng   \etal  (2004)   have   shown  that   the
semi-analytical models of Cole  \etal (2000) predict that the fraction
of `young'  galaxies decreases with  increasing halo mass up  to $\sim
10^{13}  h^{-1} \Msun$,  after which  the fraction  remains  constant. 
Although `young'  galaxies are  not necessarily the  same as  our late
types, this again seems inconsistent  with the findings reported here. 
It remains to  be seen whether this inconsistency  disappears when for
example AGN feedback is taken  into account, or whether it signals the
need for additional processes to describe type transformations.

\subsubsection{Ram pressure stripping}
\label{sec:rampres}

Whenever  a  galaxy orbits  a  hot,  gaseous  halo it  may  experience
ram-pressure  stripping (Gunn  \&  Gott 1972).   This  causes a  rapid
removal of gas,  shutting off star formation, and  transforming a late
type into  an early  type. Note, however,  that the morphology  of the
galaxy  is not  modified:  a disk  will  remain a  disk. Ram  pressure
stripping  is therefore  mainly invoked  as a  mechanism  to transform
spirals into  S0s.  Since  the latter are  typically red  and passive,
they are part of the early types in our classification scheme.

In order to  estimate how the effectiveness of  ram pressure stripping
depends on the masses of the host halo and the galaxy, consider a halo
of mass  $M$ and circular velocity  $V$.  In addition,  we assume that
the galaxy is embedded in a  subhalo of mass $m$ and circular velocity
$v$.  We  assume that both  $M$ and $m$  obey the virial  relations so
that $M  \propto V^3$ and $m  \propto v^3$.  Now consider  a disk with
surface density $\Sigma_{\rm disk}$, consisting of both stars and gas,
embedded within  $m$.  The pressure exerted  on the gas  in this disk
due to  the hot gas associated  with $M$ is $P  \propto \rho_{\rm hot}
V^2 \propto f_{\rm hot} V^2$, where $f_{\rm hot}$ is the baryonic mass
fraction of $M$  that is in a hot component  (note that all virialized
haloes have the same average  density, independent of halo mass).  The
restoring force per unit area on  the gas disk due to the self-gravity
of the  disk is $F_{\rm  res} = 2  \pi G \Sigma_{\rm  gas} \Sigma_{\rm
  star} = 2 \pi G (1 - f_{*}) f_{*} \Sigma^2_{\rm disk}$, with $f_{*}$
the disk mass fraction in stars.  To relate $\Sigma_{\rm disk}$ to the
subhalo mass $m$ we use the  disk formation models of Mo, Mao \& White
(1998),  according  to  which  $\Sigma_{\rm disk}  \propto  v  \propto
m^{1/3}$.

Ram pressure  stripping occurs  when $P >  F_{\rm res}$, which  is the
case whenever
\begin{equation}
\label{rampres}
{f_{\rm hot} \over f_{*} (1 - f_{*})} \, 
\left( {m \over M} \right)^{-2/3} > c
\end{equation}
with  $c$  some  constant.   If  we (naively)  assume  that  satellite
luminosity   is   a   reasonable   proxy   for   $m$,   we   can   use
equation~(\ref{rampres})  to predict qualitatively  how the  late type
fractions  should  scale  with  halo  mass and  halo-centric  radius.  
According  to~(\ref{rampres}), at  fixed luminosity  the  ram pressure
efficiency  scales as  $M^{2/3}$, or  even stronger  if  $f_{\rm hot}$
increases  with halo  mass as  suggested by  X-ray  measurements. This
predicts  a late  type  fraction  that decreases  with  halo mass,  as
observed.  At  fixed halo mass $M$,  however, equation (\ref{rampres})
predicts  that the  efficiency  of ram  pressure  stripping scales  as
$m^{-2/3}$.  This  implies a late  type fraction, at fixed  halo mass,
which  {\it increases}  with increasing  luminosity, which  is clearly
inconsistent with the data.  Furthermore, if ram pressure stripping is
the  main process  responsible  for the  radial  type dependence,  one
predicts the effect to be  more pronounced in more massive haloes, and
in  haloes of  fixed  mass to  be  less pronounced  for more  luminous
satellites.  Both of these predictions are inconsistent with the data,
which  shows no  luminosity  dependence  at fixed  halo  mass, and  an
equally strong radial trend for all halo masses. We therefore conclude
that ram pressure stripping can not be the dominant effect that causes
type transformations.   A similar conclusion was  recently obtained by
Goto (2005) based on a  detailed study of the velocity distribution of
galaxies within clusters.

\subsubsection{Strangulation}
\label{sec:strang}

As long  as a (central)  galaxy continues to  accrete new gas,  it can
continue to  form stars.  As  soon as it  enters a larger  system, and
becomes a satellite  galaxy, it is deprived of its  hot gas reservoir. 
This shuts  off the accretion of  new gas, so that  the star formation
rate will come  to a halt after the galaxy has  consumed (part of) its
cold  gas.  This  supply-driven  decline in  star  formation rates  of
satellite galaxies was first  suggested by Larson, Tinsley \& Caldwell
(1980), and is often called `strangulation' (Balogh, Navarro \& Morris
2000).

The main  difference between strangulation and  ram pressure stripping
is  that the  time scale  for strangulation  is much  longer  than for
stripping.  It  has been argued  that such long quenching  time scales
are  inconsistent  with  the  observation  that  the  distribution  of
H$\alpha$ equivalent widths of  starforming galaxies is independent of
environment (Balogh \etal 2004a). However, using the relations between
three different  SFR indicators, Kauffmann \etal  (2004) have actually
argued in  favor of a long time  scale ($> 1$ Gyr)  for star formation
suppression.  More detailed modeling is required to investigate these
issues in  more detail. Important  constraints may also come  from the
pronounced bimodality  in the colour magnitude  relation (e.g., Balogh
\etal 2004b; Bell \etal 2004).

Unlike  ram  pressure stripping  and  harassment,  strangulation is  a
standard  ingredient   in  most  semi-analytical   models  for  galaxy
formation (Kauffmann, White \&  Guiderdoni 1993; Diaferio \etal 2001),
where it  helps to  explain the enhanced  early type fraction  in more
massive haloes,  simply because they contain more  satellite galaxies. 
As  with  ram pressure  stripping,  however,  strangulation will  only
modify the  colours and SFRs,  but not the actual  morphologies. Thus,
while it may be an  important process to explain the enhanced fraction
of  S0  galaxies  in  dense  environments,  it  can  not  explain  the
enhancement of spheroidals.

\subsubsection{Harassment}
\label{sec:haras}

Dark matter  haloes are  populated with numerous  subhaloes of  a wide
range  of masses  (e.g.,  Gao \etal  2004;  van den  Bosch, Tormen  \&
Giocoli 2005c). A satellite galaxy embedded in one of these subhaloes,
is  subject to  frequent high  speed encounters  with  other subhaloes
(some  of  which may  not  host  a  luminous satellite  galaxy).   The
impulsive heating  due to these  numerous encounters is  termed galaxy
harassment   (Moore   \etal  1996),   and   may  cause   morphological
transformations.   In  the  tidal  approximation (Spitzer  1958),  the
amount of heating  per encounter scales as $\Delta  E \propto b^{-4}$,
with $b$ the impact parameter. To get an estimate of the total heating
due to  impulsive encounters, it is therefore  important to accurately
account   for   the  encounters   with   small   impact  parameters.   
Unfortunately, the  tidal approximation  is only valid  for relatively
large  impact  parameters (Aguilar  \&  White  1985).   This makes  it
extremely difficult to make accurate predictions regarding the scaling
of the  harassment efficiency with halo  mass. 

Nevertheless, one  point is worth making.  Galaxy  harassment is often
considered a  mechanism that  only operates in  clusters of  galaxies. 
This seems to be motivated  by the fact that clusters contain hundreds
to thousands of galaxies, very  different from groups and galaxy sized
haloes. However, in  terms of dark matter subhaloes,  the CDM paradigm
predicts  that lower mass  haloes are  simply scaled-down  versions of
cluster-sized haloes,  albeit with a relatively  small, mass dependent
normalization (e.g.,  van den Bosch  \etal 2005c).  Since  dark matter
subhaloes without a luminous satellite galaxy can also cause impulsive
heating,  galaxy harassment  is expected  to  occur in  haloes of  all
masses.

Although we  cannot make  a robust prediction  for how  the harassment
efficiency scales with  halo mass, we may use  the tidal approximation
to  estimate how  it scales  with the  mass of  the perturbed  system. 
Consider a system $s$, with  mass $m_s$, that experiences an impulsive
encounter with a perturbed $p$  of mass $m_p$.  The energy increase of
$s$ is given by
\begin{equation}
\label{enerperenc}
\Delta E = {4 \over 3} G^2 m_s {m_p^2 \over V^2} 
{\langle r_s^2 \rangle \over b^4}
\end{equation}
(Spitzer  1958), with  $b$  the impact  parameter,  $V$ the  encounter
velocity, and $\langle  r_s^2 \rangle$ the mean square  radius of $s$. 
We can express  the harassment efficiency as the  ratio of this energy
change  to the  gravitational  binding  energy of  $s$,  $W \propto  G
m_s^2/r_s$.   If we use  that $\langle  r_s^2 \rangle  \propto r_s^2$,
which  holds   as  long  as   all  systems  have  a   similar  density
distribution,  and we  assume that  the virial  relation  $m_s \propto
r_s^3$ holds, we obtain that
\begin{equation}
\label{haraseff}
\epsilon_{\rm haras} \equiv {\Delta E \over W} \propto 
\left({m_p \over V}\right)^{2} {1 \over b^4} 
\end{equation}
Note  that this is  independent of  $m_s$. If  harassment is  the main
cause of  type transformations,  and $m_s$ is  a reasonable  proxy for
satellite  luminosity, this  scaling relation  predicts that  the type
fraction should be independent of luminosity at fixed halo mass. As we
have  shown   in  Section~\ref{sec:massdep}  this   is  in  reasonable
agreement with the data, but only over the luminosity range $0.25 \lta
L/L^{*} \lta 2.5$.  However, this  argument is based on the assumption
of self-similarity. Although this  is a reasonable assumption for dark
matter subhaloes,  it does not  apply for the satellite  galaxies that
reside in these subhaloes. As shown by Moore \etal (1999), low surface
brightness (LSB) galaxies are  much more vulnerable to harassment than
high surface  brightness (HSB) galaxies  in a halo  of the same  mass. 
Since LSB galaxies have typically lower luminosities than HSB systems,
they are  expected to reside in  lower mass subhaloes,  on average. In
this case, harassment will tend to  have a bigger impact on lower mass
systems.  If harassment transforms  late type galaxies into early type
galaxies, this will result in a late-type fraction that increases with
increasing luminosity (in a halo  of fixed mass), in disagreement with
the  data. Although  clearly more  detailed studies  of the  impact of
galactic  harassment  are  needed,  these  simple  arguments  seem  to
disfavor harassment as a dominant physical process.

\subsection{Galactic Conformity}
\label{sec:conform}

In  the  standard `nature'  picture,  adopted  in all  semi-analytical
models  of galaxy  formation, the  morphology of  a central  galaxy is
related  to the  epoch  of the  last  major merger,  and  thus to  the
assembly history  of its dark  matter halo: haloes that  experienced a
recent major merger,  and thus assembled recently, are  more likely to
host  a central early  type.  Interestingly,  using a  large numerical
simulation,  Gao, Springel \&  White (2005)  have recently  shown that
haloes  {\it of  given mass}  that  assemble later  are less  strongly
biased (i.e.,  are less strongly  clustered).  If, for some  reason, a
less strongly biased region produces a larger fraction of early types,
this correlation between assembly redshift and halo bias might provide
an explanation for galactic conformity.  However, this picture has two
important  shortcomings. First  of all,  it  is well  known that  less
massive haloes are less strongly biased (e.g., Mo \& White 1996). If a
higher bias indeed results in a smaller early type fraction, one would
therefore expect an early type fraction that decreases with increasing
halo mass,  opposite to what  is observed. Secondly, Gao  \etal (2005)
have shown that the bias only depends on halo assembly time for haloes
less massive than $\sim  10^{13} h^{-1} \Msun$.  Our results, however,
indicate  that galactic  conformity  is present  in  haloes both  more
massive and less massive than this.

Alternatively, galactic  conformity might owe to  `nurture' processes. 
For example, X-ray observations show that haloes with pronounced X-ray
emission contain virtually always  an early type central galaxy (e.g.,
Ebeling, Voges \& B\"ohringer 1994; Osmond \& Ponmon 2004).  Since the
presence of X-ray emission indicates a relatively dense, hot gas halo,
conformity  might simply reflect  an enhanced  early type  fraction of
satellites due to ram pressure  stripping.  However, as we have argued
above, if  ram pressure stripping is the  dominant process responsible
for type transformations,  one would expect that, at  given halo mass,
the   early  type   fraction  decreases   with   increasing  satellite
luminosity, opposite  to what is  observed.  Alternatively, conformity
might be related  to strangulation, in which case  satellites in haloes
with  a late  type  central galaxy  need  to have  been accreted  more
recently (so that their SFRs  are not yet completely quenched).  It is
unclear,  however, why  this would  be  the case.   The final  nurture
process that  we have  discussed in this  paper, harassment,  does not
seem to provide a natural  explanation for conformity either: there is
no obvious reason why haloes with an early type central should have an
enhanced harassment rate compared to haloes of the same mass, but with
a late type central.

Clearly, galactic  conformity poses  an intriguing, new  challenge for
galaxy formation  models.  It  remains to be  seen whether  the latest
semi-analytical models  that include AGN feedback  (Croton \etal 2005)
can explain  conformity, or whether additional,  new model ingredients
are required.

\subsection{The Physical Nature of Intermediate Type Galaxies}
\label{sec:inter}

We have shown that the fraction of intermediate type galaxies is $\sim
0.2$, independent of luminosity, independent of halo mass, independent
of halo-centric  radius, and  independent of whether  the galaxy  is a
central galaxy or a satellite.  Intermediate type galaxies are defined
as galaxies  that are  `active', yet `red'  (both with respect  to the
magnitude-dependent bimodality scales). They  occupy the region in the
colour-SSFR  plane where  the early  and late  type branches  overlap. 
Therefore, it  seems natural to assume  that they consist of  a mix of
dusty  late  types  (probably  due  to an  edge-on  appearance,  which
enhances the amount of extinction) and early types with a SSFR that is
overestimated.   As discussed  in  Brinchmann \etal  (2004), the  star
formation  rate  of galaxies  with  colours  redder than  $^{0.1}(g-r)
\simeq  0.7$  are   uncertain  by  an  order  of   magnitude,  due  to
degeneracies between age, metallicity and dust.

If the intermediate types  are predominantly early (late) types, their
halo occupation  statistics should reflect  those of the  early (late)
types,  which   they  clearly  do  not.   Therefore,   if  indeed  the
intermediate types  consist of early and late  types, their fractional
contribution  must be  close to  50  percent at  all luminosities,  in
haloes  of all  masses, and  at  all halo-centric  radii.  This  seems
extremely contrived.  However, alternative explanations seem even more
implausible.  For  example, if the  intermediate types are a  class of
galaxies that are  truly distinct from early and/or  late types, it is
at least as  puzzling why they account for 20  percent of all galaxies
independent of luminosity, mass,  or radius.  Clearly, a more in-depth
investigation  regarding  the nature  of  this  class  of galaxies  is
required to provide more insight into their nature.

\section{Summary}
\label{sec:concl}

Using  the halo-based  group finder  of  Yang \etal  (2005a), we  have
constructed a large  galaxy group catalogue from the  SDSS NYU-VAGC of
Blanton \etal  (2005a).  Group (halo)  masses are determined  from the
group luminosity, which, as we have demonstrated, yields more reliable
halo masses than  using the velocity dispersion of  the group members. 
Our  catalogue also  contains  `groups' (haloes)  with  only a  single
member.   This allows us  to consider  a significantly  larger dynamic
range of  halo masses. The  final catalogue consists of  $\sim 92,000$
galaxies in $\sim 53,000$ groups  with masses $M \gta 3 \times 10^{11}
h^{-1} \Msun$. For  97 percent of these galaxies  we have obtained the
stellar masses and SFRs from the catalogues Kauffmann \etal (2003) and
Brinchmann \etal (2004), respectively.

In this first paper in  a series, we have investigated the correlation
between various galaxy properties  and halo mass.  Using the magnitude
dependent bimodality  scale in  the colour-magnitude relation  we have
split the population of galaxies into `red' and `blue' subsamples.  In
addition,  we have  used the  relation between  magnitude and  SSFR to
split  the galaxies  into  `active' and  `passive'.   The majority  of
galaxies are either `red' and `passive' (we call these early types) or
`blue' and `active'  (which we call late types).   About 20 percent of
all galaxies, however, are `red'  and `active', while only one percent
are  `blue' and  `passive'.  Except  for this  latter  minority class,
galaxies follow a tight correlation  between colour and SSFR, with two
distinct branches:  one populated  by early types,  the other  by late
types.  These two branches overlap at $^{0.1}(g-r) \sim 0.9$ and ${\rm
  log}({\rm  SSFR}/{\rm yr}^{-1})  \sim  -10.2$, where  the `red'  and
`active'  galaxies are  located.   Without further  information it  is
unclear whether these are a  physically distinct class of galaxies, or
whether they  are mainly early  types (probably with  an overestimated
SSFR) or  mainly late types  (probably edge-on disks).   Therefore, we
have provisionally called them intermediate types.

Using  our group  catalogue,  we have  investigated  the various  type
fractions as  function of halo mass, halo-centric  radius, and central
galaxy type. The main results are:

\begin{itemize}
    
\item The  early (late)  type fraction increases  (decreases) strongly
  with increasing  luminosity. This luminosity  dependence is stronger
  for    central     galaxies    than    for     satellite    galaxies
  (Section~\ref{sec:lumdep}).
 
\item  At fixed  halo mass,  the  early type  fraction increases  only
  weakly   with  increasing  luminosity.    Most  of   the  luminosity
  dependence  is only evident  at the  bright and  faint end.   In the
  regime  $0.25 \lta  L/L^*  \lta 2.5$  the  luminosity dependence  is
  insignificant.   This  holds  over  the  entire  mass  range  probed
  ($10^{12}  h^{-1} \Msun  \leq  M \leq  10^{15}  h^{-1} \Msun$),  and
  implies  that  halo  mass  is  more important  for  determining  the
  properties  of a galaxy  than is  galaxy luminosity.   A significant
  part of the  strong luminosity dependence is simply  a reflection of
  the fact that  more luminous galaxies reside in  more massive haloes
  (Section~\ref{sec:massdep}).
    
\item At  fixed luminosity, the  early (late) type  fraction increases
  (decreases) with  increasing halo  mass.  Most importantly,  we find
  that this  mass dependence is smooth  and that it  persists over the
  entire mass range  probed: {\it there is no break  or feature at any
    mass  scale}.  This  differs from  previous work.   In particular,
  various studies  have found that the  environment dependence becomes
  weaker,  or  completely  vanishes,  below a  characteristic  density
  scale.  This has been  interpreted as indicating that group-specific
  processes are  the dominant cause of type  transformations.  We have
  argued, however, that this  characteristic scale merely reflects the
  scale  at  which  the  physical  meaning of  the  density  estimator
  transits from  a local  density ($R <  R_{\rm vir}$) estimator  to a
  global,  large scale  density ($R  > R_{\rm  vir}$)  estimator.  Our
  results, based  on halo masses,  find no indication  whatsoever that
  group-  and/or cluster-specific  processes play  a dominant  role in
  type transitions (Section~\ref{sec:massdep}).
  
\item  The  early  (late)  type fraction  decreases  (increases)  with
  increasing halo-centric  radius.  Contrary to  previous studies, who
  found no radius  dependency in haloes with $M  \lta 10^{13.5} h^{-1}
  \Msun$, we  find a self-similar  dependence in haloes of  all masses
  probed ($10^{12}  h^{-1} \Msun \leq  M \leq 10^{15} h^{-1}  \Msun$). 
  This  discrepancy is  most  likely  due to  the  fact that  previous
  studies   included  the   central   galaxies  and   were  based   on
  significantly smaller samples (Section~\ref{sec:raddep}).
  
\item The intermediate type fraction is $\sim 20$ percent, independent
  of luminosity, independent of halo mass, independent of halo-centric
  radius, and independent of whether the galaxy is a central galaxy or
  a  satellite  galaxy.   Probably  the easiest  explanation  is  that
  intermediates  consist of  an equal  mix of  early and  late  types. 
  Although consistent with the fact that intermediate types lie in the
  region  in the  colour-SSFR  plane  where the  early  and late  type
  branches overlap,  it is extremely puzzling that  the fractional mix
  does not scale with luminosity, halo mass, or halo-centric radius. A
  more in-depth  study is required  to investigate the nature  of this
  class  of   galaxies  in  more   detail  (Sections~\ref{sec:massdep}
  and~\ref{sec:raddep}).
  
\item  The properties of  a satellite  galaxy are  strongly correlated
  with those of its central  galaxy. In particular, we have shown that
  the early type fraction of  satellites is significantly higher for a
  halo with an  early type central galaxy than for a  halo {\it of the
    same mass} but with a  late type central galaxy.  This phenomenon,
  which we  call `galactic  conformity', is present  in haloes  of all
  masses     and     for     satellites    of     all     luminosities
  (Section~\ref{sec:typedep}).
  
\item The median physical  properties of late, early, and intermediate
  type galaxies of a given luminosity  do not depend on halo mass. The
  relative  fractions of  these types,  however, do.   Since different
  galaxy  types  have  different  median properties,  this  halo  mass
  dependence of  the type fractions  causes a halo mass  dependence of
  the    median   properties   of    the   full    galaxy   population
  (Section~\ref{sec:propmass}).

\end{itemize}

We have discussed  the possible implication of these  findings for our
understanding of galaxy formation  and evolution. Using simple scaling
arguments, we have argued that  both ram pressure stripping and galaxy
harassment  are  not  the   major  processes  responsible  for  galaxy
transformations, as they predict an increasing late type fraction with
increasing luminosity  in haloes  of fixed mass,  opposite to  what is
observed.  We therefore suggest  that merger history and strangulation
(i.e., the quenching  of star formation as soon as  a galaxy becomes a
satellite galaxy) are the main ingredients required to predict whether
a galaxy ends up as an early or a late type galaxy.

This conclusion, however, is still extremely speculative. For example,
it still needs to be seen, whether the semi-analytical models that use
strangulation  and the  merger history  to predict  galaxy  types, are
indeed  consistent  with the  various  observational trends  presented
here. In particular, we have  argued that galactic conformity poses an
intriguing new  challenge for  galaxy formation models.   Although the
correlations between  galaxy properties  and halo mass  presented here
provide an interesting testbed for galaxy formation models, a definite
explanation for the origin of the bimodality of galaxy properties will
most likely have to await a similar analysis as performed here, but at
different epochs  (i.e., different redshifts).  It  is reassuring that
promising work  in this  direction is already  under way  (e.g., Gerke
\etal 2005; Cooper \etal 2005).

\section*{Acknowledgements}

We  are extremely  grateful to  Michael Blanton,  Guinevere Kauffmann,
Jarle Brinchmann and all other people within the SDSS collaboration for
producing a wonderful data set and for making their various catalogues
publicly available. Michael Blanton  is also acknowledged for his help
with  the NYU-VAGC.   FvdB acknowledges  useful discussions  with Eric
Bell, Aaron Dutton, and Anna Pasquali.



\appendix

\section[]{The Group-Finding Algorithm}
\label{sec:AppA}

The group finder, used  in Section~\ref{sec:groupcat} to construct our
SDSS  group  catalogue uses  some  virial  properties  of dark  matter
haloes.   Throughout  this  paper  we  define dark  matter  haloes  as
virialized  structures with  a mean  over-density of  $180$ and  a NFW
(Navarro, Frenk \& White 1997) density distribution:
\begin{equation}
\label{NFW}
\rho(r) = \frac{\bar{\delta}\bar{\rho}}{(r/r_{\rm s})(1+r/r_{\rm  s})^{2}}
\end{equation}
Here  $r_s$ is a  characteristic radius,  $\bar{\rho}$ is  the average
density  of  the  Universe,  and  $\bar{\delta}$  is  a  dimensionless
amplitude which  can be expressed  in terms of the  halo concentration
parameter $c=r_{180}/r_s$ as
\begin{equation}
\label{overdensity}
\bar{\delta} = {180 \over 3} \, {c^{3} \over {\rm ln}(1+c) - c/(1+c)},
\end{equation}
with  $r_{180}$  the radius  within  which  the  halo has  an  average
over-density of  $180$.  We  use the relation  given by  Bullock \etal
(2001) to compute $c$ as function  of halo mass, properly converted to
our definition of halo mass.

Our group finder consists of the following steps:

{\bf Step 1:} We combine two different methods to identify the centers
of potential  groups. First  we use the  traditional FOF  algorithm to
assign galaxies to groups.  Since we are working in redshift space, we
separately define  linking lengths along the line  of sight ($\ell_z$)
and in the transverse direction ($\ell_p$).  Since the purpose here is
only to  identify the group  centers, we use relatively  small linking
lengths:  $\ell_z=0.3$ and $\ell_p=0.05$,  both in  units of  the mean
separation of  galaxies. Note that  for an apparent  magnitude limited
survey  the mean  separation of  galaxies is  a function  of redshift,
which  we take  into account.   The geometrical,  luminosity weighted,
centers of  all FOF groups thus  identified with two  galaxies or more
are  considered  as  centers  of  potential groups.   Next,  from  all
galaxies  not yet  linked  together  by these  FOF  groups, we  select
bright, relatively isolated galaxies  which we also associate with the
centers of  potential groups.  Following an approach  similar to McKay
\etal (2002),  Prada \etal (2003)  and Brainerd \& Specian  (2003), we
identify a galaxy as `central', and  thus as the center of a potential
group,  when  it is  the  brightest galaxy  in  a  cylinder of  radius
$1\mpch$ and a velocity depth of $500\kms$.

{\bf Step 2:} We estimate the luminosity of a selected potential group
using
\begin{equation}
\label{Lgroup}
L_{\rm group} = \sum_i {L_i \over c_i} 
\end{equation}
Here $L_i$ is the $^{0.1}r$-band luminosity of the $i^{\rm th}$ galaxy
in  the  group  and  $c_i$  is the  SDSS  survey-completeness  at  the
corresponding   location.  The   {\it  total}   group   luminosity  is
approximated by
\begin{equation}
\label{Ltotl}
L_{\rm total} = L_{\rm group} \frac{\int_0^{\infty} \Phi(L) \, L \, dL}
{\int_{L_{\rm lim}}^{\infty} \Phi(L) \, L \, dL}\,,
\end{equation}
where $L_{\rm lim}$ is the minimum  luminosity of a galaxy that can be
observed at  the redshift  of the group,  and $\Phi(L)$ is  the galaxy
luminosity function  in the $^{0.1}r$-band,  which we model  using the
Schechter function fit of Blanton \etal (2003a).

{\bf  Step  3:}  From  $L_{\rm  total}$  and a  model  for  the  group
mass-to-light ratio  (see below), we  compute an estimate of  the halo
mass  associated with  the  group in  consideration.   From this  mass
estimate we compute  the halo radius $r_{\rm 180}$,  the virial radius
$r_{\rm  vir}$\footnote{The virial  radius  is defined  as the  radius
  inside of which the average  density is $\Delta_{\rm vir}$ times the
  critical density,  with $\Delta_{\rm vir}$ given by  Bryan \& Norman
  (1998)}, and the line-of-sight velocity dispersion $\sigma$. For the
latter we use
\begin{equation}
\label{veldispfit}
\sigma = 428.0 \kms 
\left( {M_{180} \over 10^{14} h^{-1} \Msun}\right)^{0.3244}
\end{equation}
This  fitting  function  accurately  describes  the  relation  between
$M_{180}$ and the mass-weighted one-dimension velocity dispersion (see
equation (14) in van den Bosch \etal 2004).

{\bf  Step  4:} Using  the  sizes,  masses,  velocity dispersions  and
centers of the  groups thus obtained, we now  assign group memberships
to  all  galaxies  in  the  survey. We  assume  that  the  phase-space
distribution of  galaxies follows that  of the dark matter  particles. 
In that case the number density contrast of galaxies in redshift space
around the  group center  (which we associate  with the center  of the
dark matter halo) at redshift $z_{\rm group}$ is given by
\begin{equation}
P_M(R,\Delta z) = {H_0\over c} {\Sigma(R)\over {\bar \rho}} p(\Delta z) \,,
\end{equation}
Here $\Delta z  = z - z_{\rm group}$ and  $\Sigma(R)$ is the projected
surface density of a (spherical) NFW halo:
\begin{equation}
\Sigma(R)= 2~r_s~\bar{\delta}~{\bar \rho}~{f(R/r_c)}\,,
\end{equation}
with 
\begin{equation}\label{eq:fx}
f(x) = \left\{
\begin{array}{lll}
\frac{1}{x^{2}-1}\left(1-\frac{{\ln
{\frac{1+\sqrt{1-x^2}}{x}}}}{\sqrt{1-x^{2}}}\right)
 & \mbox{if $x<1$} \\
\frac{1}{3}
 & \mbox{if $x=1$} \\
\frac{1}{x^{2}-1}\left(1-\frac{{\rm
atan}\sqrt{x^2-1}}{\sqrt{x^{2}-1}}\right)
 & \mbox{if $x>1$}
\end{array} \right.\,.
\end{equation}
The  function $p(\Delta  z){\rm d}\Delta  z $  describes  the redshift
distribution of galaxies within the halo for which we adopt a Gaussian
form
\begin{equation}
p(\Delta z)= {1 \over \sqrt{2\pi}} {c \over \sigma (1+z_{\rm group})} 
\exp \left [ \frac {-(c\Delta z)^2}
{2\sigma^2(1+z_{\rm group})^2}\right ] \,,
\end{equation}
where $\sigma$ is the rest-frame velocity dispersion.

Thus  defined,  $P_M(R,\Delta  z)$  is the  three-dimensional  density
contrast  in redshift  space.  In  order  to decide  whether a  galaxy
should be assigned  to a particular group we  proceed as follows.  For
each galaxy  we loop  over all groups,  and compute  the corresponding
distance $(R,\Delta z)$ between galaxy  and group center.  Here $R$ is
the projected distance at the  redshift of the group. If $P_M(R,\Delta
z)  \ge B$,  with $B$  an appropriately  chosen background  level (see
below),  the galaxy  is assigned  to  the group.  If a  galaxy can  be
assigned to more than one group,  it is only assigned to the group for
which  $P_M(R,\Delta  z)$ has  the  highest  value.   Finally, if  all
members of  two groups can be  assigned to one group  according to the
above criterion, the two groups are merged into a single group.

{\bf Step 5:}  Using the group members thus  selected we recompute the
geometrical, luminosity  weighted group center and go back  to Step 2,
iterating  until there  is no  further  change in  the memberships  of
groups.  Note that, unlike with the traditional FOF method, this group
finder also identifies groups with only one member.

The group finding algorithm thus defined requires an assumed $M/L_{\rm
  group}$, possibly  as function  of halo mass  $M$, and has  one free
parameter,  namely the  background level  $B$.  In  this paper  we use
$B=10$,  which  corresponds  roughly  to  the  redshift-space  density
contrast at the edge of a halo (see YMBJ). As shown in YMBJ, the group
catalogue is not  very sensitive to the exact value  of $B$ used.  For
$M/L_{\rm group}$ we use  the average mass-to-light ratios as function
of halo mass obtained by van  den Bosch, Yang \& Mo (2003; their model
D).  Since  mass-to-light ratio  corresponds to the  photometric $b_J$
band,  we  compute  $L_{\rm group}$  in  both  the  $g$ and  $r$  band
($k$-corrected to  $z=0$), which  we convert to  the $b_J$  band using
$b_J =  g + 0.155  + 0.15238 *  (g-r)$ (Fukugita \etal 1996;  Blair \&
Gilmore 1982). Detailed tests in YMBJ have shown that the completeness
and  contamination  levels  of   our  group  catalogue  are  extremely
insensitive to  the exact values  of $M/L$ assumed.  We  have verified
that very  significant changes in this assumption  have no significant
effect of any of the results presented in this paper.  This is easy to
understand; even if our estimate for  $M/L$ is wrong by a factor of 3,
the  implied radius and  velocity dispersion,  used in  the membership
determination, are only off by 44 percent.

\begin{figure*}
\centerline{\psfig{figure=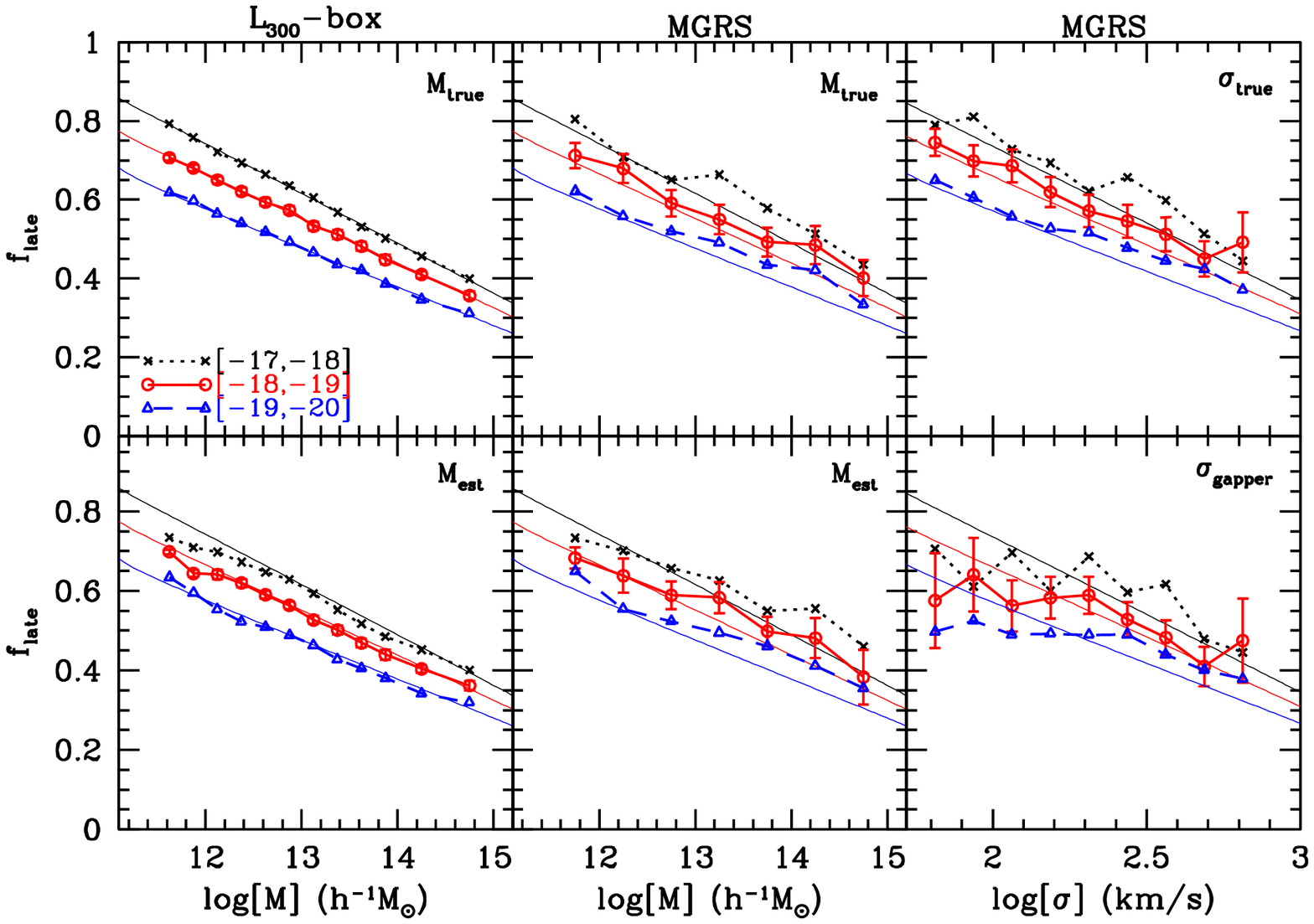,width=0.9\hdsize}}
\caption{Late type fractions as function of halo mass and velocity
  dispersion. The left-hand panels  show the results obtained directly
  from the  $L_{300}$ simulation  box, using both  the true  halo mass
  ($M_{\rm true}$,  upper panel) and the estimated  halo mass ($M_{\rm
    est}$, lower  panel). Panels in  the middle column show  the same,
  but this  time based on a  group catalogue extracted from  the MGRS. 
  Right-hand panels show the same results but this time as function of
  velocity dispersion,  rather than halo mass.  Results  are shown for
  three magnitude-limited  samples; the  values in square  brackets in
  the upper left-hand panel indicate the range of $M_{b_J}-5 {\rm log}
  h$  used.  The  thin  lines in  each  panel correspond  to the  true
  underlying fractions as specified by  the CLF.  For clarity, we only
  plot (Poissonian) errorbars for the  sample with $-18 \geq M_{b_J} -
  5 {\rm log} h > -19$.  See text for a detailed discussion.}
\label{fig:mock}
\end{figure*}

\section[]{Testing the Robustness of our Results with Mock Surveys}
\label{sec:AppB}

In order to address the robustness  of our results we construct a mock
galaxy redshift  survey (hereafter MGRS), which we  analyze in exactly
the  same  way  as  the  data  described in  the  previous  sections.  
Even though our data is based on  the SDSS, we follow Yang \etal (2005a)
and construct a mock version of the 2dFGRS.  There are two reasons for
this.  First of  all,  we do  not  yet have  accurate  models for  the
conditional luminosity  function of  SDSS galaxies, which  is required
for  the construction of  reliable MGRSs  (see below).   Secondly, our
mock versions  of the  2dFGRS have been  well tested and  are accurate
representations of  the actual  2dFGRS (see Yang  \etal 2005a;  van den
Bosch \etal  2005a,b). Since the main  purpose of this  exercise is to
test the  methodologies used in this  paper, the use of  a mock 2dFGRS
rather than a  mock SDSS should not make  a significant difference. If
anything, since the SDSS sample  used here is somewhat larger than the
2dFGRS, and  since the redshift errors are  substantially smaller, our
results  regarding  the reliability  and  robustness  of the  analysis
should be considered conservative.

The MGRS is constructed by populating dark matter haloes with galaxies
of different  luminosities. The distribution of dark  matter haloes is
obtained from a  set of large $N$-body simulations  (dark matter only)
for  a $\Lambda$CDM  `concordance'  cosmology with  $\Omega_m =  0.3$,
$\Omega_{\Lambda}=0.7$, $h=0.7$ and  $\sigma_8=0.9$.  In this paper we
use two simulations with $N=512^3$ particles each, which are described
in more detail in Jing  \& Suto (2002).  The simulations have periodic
boundary  conditions and box  sizes of  $L_{\rm box}=100  h^{-1} \Mpc$
(hereafter  $L_{100}$) and  $L_{\rm box}=300  h^{-1}  \Mpc$ (hereafter
$L_{300}$). Dark  matter haloes are identified using  the standard FOF
algorithm with a linking length of $0.2$ times the mean inter-particle
separation.  

To  populate  these dark  matter  haloes  with  galaxies of  different
luminosities and  different types,  we use the  conditional luminosity
function (hereafter  CLF), $\Phi(L \vert M)$, which  gives the average
number of  galaxies of luminosity $L$  that resides in a  halo of mass
$M$ (Yang, Mo \& van den  Bosch 2003). The sample of galaxies is split
in `early types' and `late types' using a probability function $P_{\rm
  late}(L,M)$ (see van den Bosch,  Yang \& Mo 2003).  Details of these
models are  not important for what  follows, but we do  point out that
these models  accurately fit the luminosity  functions (Madgwick \etal
2002) and the  correlation lengths as function  of luminosity (Norberg
\etal 2002) for both galaxy types.

Having populated  the various simulation boxes with  galaxies we first
proceed as  follows.  In each  halo we count  the number of  early and
late type galaxies in a given magnitude range, and compute the average
late  type fraction  as function  of halo  mass. The  results  for the
$L_{300}$  box are  shown,  for three  different  magnitude ranges  as
indicated,  in  the   upper  left-hand  panel  of  Fig.~\ref{fig:mock}
(symbols  connected  by  thick   lines).   The  thin,  lines  are  the
theoretical predictions corresponding to the input CLF, given by
\begin{equation}
\label{CLFpred}
f_{\rm late}(M) = {\int_{L_{\rm min}}^{L_{\rm max}} P_{\rm late}(L,M) \, 
   \Phi(L \vert M) \, \rmd L \over \int_{L_{\rm min}}^{L_{\rm max}} 
   \Phi(L \vert M) \, \rmd L}\,.
\end{equation}
Here  $L_{\rm  min}$  and  $L_{\rm  max}$ are  the  luminosities  that
correspond to  the magnitude limits.  Not  surprisingly, the late type
fractions derived are in  excellent agreement with these input values;
this figure is  just to illustrate that the  box contains a sufficient
number of haloes so that the Poisson errors are negligibly small.

The lower  left-hand panel shows  the results obtained when  using the
estimated halo mass,  rather than the true halo  mass. Halo masses are
estimated    using    a   similar    procedure    as   described    in
Section~\ref{sec:groupmass}: for each  halo we determine $L_{18}$, the
total luminosity  of all halo galaxies  with $M_{b_J} - 5  {\rm log} h
\leq -18$, and we compute  the number density, $n_{+}$, of haloes with
$L_{18}$ larger than that of  the halo considered. Using the halo mass
function of Sheth, Mo \& Tormen (2001), we determine the corresponding
halo mass  by finding the  mass for which  the number density  of more
massive   haloes    is   equal   to   $n_{+}$.     As   discussed   in
Section~\ref{sec:groupmass} this  method thus assigns  masses based on
the $L_{18}$  rank order of the  haloes. As is evident  from the lower
left-hand panel of Fig.~\ref{fig:mock},  this method of assigning halo
masses results  in small, systematic  errors in the derived  late type
fractions for haloes with $M \lta 2\times 10^{12}h^{-1} \Msun$, in the
sense that the luminosity-dependence  is underestimated.  This owes to
the fact  that the  luminosities themselves are  used to  estimate the
halo masses.  For more  massive haloes, however, the resulting $f_{\rm
  late}(L,M)$ is  virtually indistinguishable from the  true relation. 
This demonstrates  that our method  of assigning halo masses  does not
introduce  any  systematic  errors   in  the  mass  and/or  luminosity
dependence  of  the galaxy  types  for  haloes  with $M  \gta  2\times
10^{12}h^{-1} \Msun$. We emphasize  that the relation between $L_{18}$
and $M$ in the MGRS has a realistic amount of scatter.
 
The above  test, however, is idealized.  In reality we  have to select
haloes using  a group finder applied  to a redshift  survey. Since the
survey  suffers  from   observational  biases  and  peculiar  velocity
distortions,  and  since the  group  finder  unavoidably suffers  from
interlopers and incompleteness effects,  a more realistic check of our
methodology  requires  a comparison  with  a  proper  MGRS. Using  the
$L_{100}$ and  $L_{300}$ simulation boxes described above  we create a
large virtual universe.  We follow Yang \etal (2005a) and replicate the
$L_{300}$ box on a $4 \times  4 \times 4$ grid.  The central $2 \times
2 \times 2$  boxes, are replaced by  a stack of $6 \times  6 \times 6$
$L_{100}$ boxes, and the virtual observer is placed at the center (see
Fig.~11  in  Yang \etal  2005a).   This  stacking geometry  circumvents
incompleteness problems  in the mock  survey due to  insufficient mass
resolution of  the $L_{300}$ simulations,  and allows us to  reach the
desired  depth  of $z_{\rm  max}=0.20$  in  all  directions.  Next  we
construct a mock  2dFGRS using the following steps  (see van den Bosch
\etal 2005a for details):

\begin{enumerate}
  
\item We  define a $(\alpha,\delta)$-coordinate frame  with respect to
  the virtual observer at the center of the stack of simulation boxes,
  and remove all galaxies that are not located in the areas equivalent
  to those of the 2dFGRS.
  
\item For each  galaxy we compute the apparent  magnitude according to
  its luminosity  and distance, to  which we add  a rms error  of 0.15
  mag. 
  
\item For each galaxy we compute the redshift as `seen' by the virtual
  observer.   We take  the observational  velocity  uncertainties into
  account  by   adding  a  random  velocity  drawn   from  a  Gaussian
  distribution with dispersion $85\kms$.
  
\item To  take  account  of  the  position-  and  magnitude-dependent
  completeness of the 2dFGRS, we randomly sample each galaxy using the
  completeness masks provided by the 2dFGRS team.
    
\item   We  also   take   account  of   the  fiber-collision   induced
  incompleteness as  well as the incompleteness due  to image
  blending. 

\end{enumerate}

As shown  in Yang \etal (2005a)  and van den Bosch  \etal (2005a), this
procedure  results in  a mock  2dFGRS that  accurately mimics  all the
various  incompleteness  effects, allowing  for  a direct,  one-to-one
comparison with the true 2dFGRS.

Next  we apply  the YMBJ  halo-based group  finder to  this  MGRS, and
compute the late type fraction as function of halo mass using both the
true halo  masses (defined as the  true halo mass  associated with the
brightest  group member)  and  the estimated  halo  masses (using  the
$L_{18}$ ranking  method described above).   The results are  shown in
the panels  in the middle column of  Fig.~\ref{fig:mock}.  Since there
are much fewer galaxies/haloes involved  than in the case shown in the
left-hand  panels, and  since  the group-finder  is  not perfect,  the
results are  significantly more  noisy.  Nevertheless, when  using the
true  halo  masses, the  resulting  late type  fractions  are in  good
agreement with  the input  values (eq.~[\ref{CLFpred}]), except  for a
small, systematic  overestimate at the massive end  due to interlopers
and incompleteness  effects.  When the estimated halo  masses are used
instead, one again notices a small but systematic underestimate of the
luminosity dependence of $f_{\rm late}(L,M)$ for haloes with $M \lta 2
\times 10^{12}  h^{-1} \Msun$.  For  more massive haloes,  the results
are  very comparable  to those  based on  the true  halo  masses. This
indicates  that  our  group   finder  allows  for  a  fairly  accurate
determination  of  $f_{\rm  late}(L,M)$.  In  particular,  the  method
accurately recovers the luminosity  dependence.  (at least for $M \lta
2 \times  10^{12} h^{-1}  \Msun$). Recall that  since the  SDSS sample
used in this  paper is larger than the 2dFGRS,  and since the redshift
errors  in the  SDSS  are  significantly smaller  than  in the  2dFGRS
(resulting in  smaller interloper  fractions), we may  actually expect
the SDSS results  presented in the previous section  to be more robust
than the MGRS results shown here.

Finally,  the  panels  on  the  right-hand side  show  the  late  type
fractions obtained from  the MGRS as function of  velocity dispersion. 
In the upper right-hand panel we plot the fractions as function of the
true  velocity  dispersion,  which  is  the  one-dimensional  velocity
dispersion of the dark matter particles corresponding to the halo that
hosts the brightest group galaxy. As expected, these results look very
similar to those in the upper panel in the middle column. In the lower
right-hand panel, however,  we plot $f_{\rm late}$ as  function of the
velocity dispersion  of the group  members, measured using  the gapper
estimator, which is insensitive  to outliers (Beers, Flynn \& Gebhardt
1990; see  YMBJ for  our implementation).  Only  groups with  at least
three members  are taken into  account.  This time, the  dependence of
$f_{\rm late}$  on the halo  velocity dispersion is much  flatter than
for the input model.   Especially for haloes with $\sigma_{\rm gapper}
\lta  160 \kms$  (corresponding to  $M  \lta 5  \times 10^{12}  h^{-1}
\Msun$),  the late  type fractions  are significantly  underestimated. 
This demonstrates that our  mass estimates based on the $L_{18}$-group
ranking are more reliable than those based on the velocity dispersion,
especially for low mass haloes.


\begin{thebibliography}{}

\bibitem[]{Aba04}
Abazajian, K. et al, 2004, \aj, 128, 502

\bibitem[]{Agu85}
Aguilar L.A., White S.D.M., 1985, \apj , 295, 374

\bibitem[]{Bal04}
Baldry I.K., Glazebrook K., Brinkmann J., Ivezi\'{c} \v{Z}., Lupton R.H.,
Nichol R.C., Szalay A.S., 2004, \apj, 600, 681

\bibitem[]{Bal97} 
Balogh M.L., Morris S.L., Yee H.K.C., Carlberg R.G.,
  Ellingson E., 1997, \apj , 488, L75

\bibitem[]{Bal99} 
Balogh M.L., Morris S.L., Yee H.K.C., Carlberg R.G.,
  Ellingson E., 1999, \apj , 527, 54

\bibitem[]{Bal00}
Balogh M.L., Navarro J.F., Morris S.L., 2000, \apj , 540, 113

\bibitem[]{Bal04a}
Balogh M.L., et al., 2004a, \mnras, 348, 1355

\bibitem[]{Bal04b}
Balogh M.L., Baldry I.K., Nichol R., Miller C., Bower R., Glazebrook
K., 2004b, \apj, 615, L101

\bibitem[]{Bee90}
Beers T.C., Flynn K., Gebhardt K., 1990, \aj , 100, 32

\bibitem[]{Bel04}
Bell E.F., Wolf C., Meisenheimer K., Rix H.-W., Borch A., Dye
S. Kleinheinrich M., Wisotzki L., McInthosh D.H., 2004, \apj, 608, 752

\bibitem[]{Ber03}
Berlind A.A., Weinberg D.H., Benson A.J., Baugh C.M. Cole S., Dav\'{e}
R., Frenk C. S., Jenkins A., Katz N., Lacey C.G.,  2003, \apj, 593, 1

\bibitem[]{Biv02}
Biviano A., Katgert P., Thomas T., Adami C., 2002, \aap , 387, 8

\bibitem[]{Bla82}
Blair M., Gilmore G., 1982, PASP, 94, 741

\bibitem[]{LF-paper1}
Blanton M. R. \etal, 2001, \aj, 121, 2358

\bibitem[]{LF-paper2}
Blanton M. R. \etal, 2003a, \apj, 592, 819

\bibitem[]{bimodality-paper}
Blanton M. R. \etal, 2003b, \apj, 594, 186

\bibitem[]{k-corr paper}
Blanton M. R. \etal, 2003c, \aj, 125, 2348

\bibitem[] {Bla04} 
Blanton M.R.,  Eisenstein D., Hogg D.W., Zehavi I.,
  2004, preprint (astro-ph/0411037)

\bibitem[]{NYUVAGC}
Blanton M.R. \etal, 2005a, \aj, 129, 2562

\bibitem[]{Environ} 
Blanton M.R.,  Eisenstein D., Hogg D.W., Schlegel D.J., Brinkman J.,
  2005b, \apj, 629, 143

\bibitem[]{Bra03}
Brainerd T.G., Specian M.A., 2003, \apj, 593, L7

\bibitem[]{Bri04}
Brinchmann J., Charlot S., White S.D.M, Tremonti C., Kauffmann G.,
Heckman T., Brinkmann J.,  2004, \mnras, 353, 713

\bibitem[]{Bry98}
Bryan G., Norman M., 1998, \apj , 495, 80

\bibitem[]{Bul01}
Bullock J.S., Kolatt T.S., Sigad Y., Somerville R.S., Kravtsov A.V.,
Klypin A.A., Primack J.R., Dekel A., 2001, \mnras, 321, 559

\bibitem[]{Byr90}
Byrd G., Valtonen M., 1990, \apj , 350, 89

\bibitem[]{Col00}
Cole S., Lacey C.G., Baugh C.M., Frenk C.S., 2000, \mnras, 319, 168

\bibitem[]{Coll01}
Colless M., The 2dFGRS team, 2001, \mnras , 328, 1039

\bibitem[]{Col05}
Collister A.A., Lahav O., 2005, \mnras , 361, 415

\bibitem[]{Cop05}
Cooper M.C., Newman J.A., Madgwick D.S., Gerke B.F., Yan R., Davis M.,
2005, preprint (astro-ph/0506518)

\bibitem[]{Coo02}
Cooray A., Sheth R., 2002, Phys. Rep. 372, 1

\bibitem[]{Coo05}
Cooray A., 2005, preprint (astro-ph/0505421)

\bibitem[]{CM05}
Cooray A., Milosavljevi\'c M., 2005, \apj , 627, L89

\bibitem[]{Cro05}
Croton D.J., et al., 2005, \mnras, 356, 1155

\bibitem[]{Dep04}
De Propris R., et al., 2004, \mnras, 351, 125D

\bibitem[]{Dia01}
Diaferio A., Kauffmann G., Balogh M.L, White S.D.M, Schade D.,
Ellingson E., 2001, \mnras, 323, 

\bibitem[]{Dom01}
Dom\'{i}nguez M.J., Muriel H., Lambas D.G., 2001, \aj , 121, 1266

\bibitem[]{Dom02}
Dom\'{i}nguez M.J., Zandivarez A.A., Mart\'{i}nez H.J., Merch\'{a}n
M.E., Muriel H., Lambas D.G., 2002, \apj, 335, 825

\bibitem[]{Dre80}
Dressler A., 1980, \apj, 236, 351

\bibitem[]{Ebe94}
Ebeling H., Voges W., B\"ohringer H., 1994, \apj , 436, 44

\bibitem[]{Efs92}
Efstathiou G., 1992, \mnras , 256, 43

\bibitem[]{Eke04a}
Eke V.R.\& The 2dFGS team, 2004a, \mnras, 348, 866

\bibitem[]{Eke04b}
Eke V.R.\& The 2dFGS team, 2004b, \mnras, 355, 769

\bibitem[]{Evr90}
Evrard A.E., Silk J., Szalay A.S., 1990, \apj , 365, 13

\bibitem[]{Fab05}
Faber S.M., et al., 2005, preprint (astro-ph/0506044)

\bibitem[]{Fuk96}
Fukugita M., Ichikawa T., Gunn J.E., Doi M., Shimasaku K., Schneider D.P.,
1996, \aj , 111, 1748

\bibitem[]{Gao04}
Gao L., White S.D.M., Jenkins A., Stoehr F., Springel V., 2004,
\mnras , 355, 819

\bibitem[]{Gao05}
Gao L., Springel V., White S.D.M., 2005, preprint (astro-ph/0506510)

\bibitem[]{Gel83}
Geller M.J., Huchra J.P., 1983, \apjs , 47, 139

\bibitem[]{Ger04}
Gerke B.F. et al., 2005, \apj , 625, 6 

\bibitem[]{Ghi98}
Ghigna S., Moore B., Governato F., Lake G., Quinn T., Stadel J., 1998,
\mnras , 300, 146

\bibitem[]{Gir03}
Girardi M., Rigoni E., Mardirossian F., Mezzetti M., 2003, \aap, 406,
403G

\bibitem[]{Gom03}
G\'{o}mez P.L., et al., 2003, \apj, 584, 210

\bibitem[]{Got03}
Goto T., Yamauchi C., Fujita Y., Okamura S., Sekiguchi M., Smail I.,
Bernardi M., Gomez P.L., 2003, \mnras, 346, 601

\bibitem[]{Got04}
Goto T., Masafumi Y., Tanaka M., Okamura S., 2004, \mnras, 348, 515

\bibitem[]{Got05}
Goto T., 2005, \mnras, 356, L6

\bibitem[]{Gun72}
Gunn J.E., Gott J.R., 1972, \apj , 176, 1

\bibitem[]{Has98}
Hashimoto Z., Oemler A., Lin H., Tucker D.L., 1998, \apj, 499, 589

\bibitem[]{Has99}
Hashimoto Z., Oemler A., 1999, \apj, 510, 609

\bibitem[]{Hic84} 
Hickson P., Ninkov Z., Huchra J., Mamon G., 1984, in
  Clusters and Groups of  Galaxies, eds. F. Mardirossian, G. Giuricin,
  M. Mezzetti, Reidel, Dordrecht, Holland, p. 367

\bibitem[]{Hog03}
Hogg D.W., et al., 2003, \apj , 585, L5

\bibitem[]{Hog04}
Hogg D.W., et al., 2004, \apj , 601, L29

\bibitem[]{Hoy05}
Hoyle F., Rojas R.R., Vogeley M.S., Brinkmann J., 2005, \apj , 620,
618

\bibitem[]{Huc82}
Huchra J.P., Geller M.J., 1982, \apj , 257, 423

\bibitem[]{JiS02}
Jing Y.P., Suto, Y., 2002, \apj , 574, 538

\bibitem[]{Kau93}
Kauffmann G., White S.D.M, Guiderdoni B., 1993, \mnras,
264, 201

\bibitem[]{Kau03}
Kauffmann G., \etal, 2003, \mnras, 341, 33

\bibitem[]{Kau04}
Kauffmann G., White S.D.M, Heckman T.M., M\'{e}nard, Brinchmann J.,
Charlot S., Tremonti C., Brinkmann J.,  2004, \mnras, 353, 713

\bibitem[]{Kel05}
Kelm B., Focardi P., Sorrentino G., 2005, preprint (astro-ph/0506531)

\bibitem[]{Kep99}
Kepner J., Fan X., Bahcall N., Gunn J., Lupton R., Xu G., 1999, ApJ,
517, 78

\bibitem[]{Kim02}
Kim R.J.S. et al, 2002, AJ, 123, 20

\bibitem[]{Koc03}
Kochanek C.S., White M., Huchra J., Macri L., Jarrett T.H., Schneider
S.E., Mader J., 2003, \apj, 585, 161

\bibitem[]{Kue05}
Kuehn F., Ryden B.S., 2005, preprint (astro-ph/0508337)

\bibitem[]{Lar80}
Larson R.B., Tinsley B.M., Caldwell C.N., 1980, \apj , 237, 692

\bibitem[]{Lem99}
Lemson G. \& Kauffmann G., 1999, \mnras, 302, 111

\bibitem[]{Lew02}
Lewis I, et al., 2002, \mnras, 334, 673

\bibitem[]{Lov92}
Loveday J., Peterson B.A., Efstathiou G., Maddox S.J., 1992, \aap, 390, 338

\bibitem[]{Mad02}
Madgwick D.S. \& the 2dFGRS team, 2002, \mnras, 333, 133

\bibitem[]{Mag03}
Magliocchetti M. \& Porciani C., 2003, \mnras, 345, 186

\bibitem[]{Mar02}
Mart\'{i}nez H.J., Zandivarez A., Dom\'{i}nguez M., Merch\'{a}n M.E.,
Lambas D.G., 2002, \mnras, 333, L31

\bibitem[]{Mar97}
Marzke R.O., da Costa L.N., 1997, \aj, 113, 185

\bibitem[]{Mar98}
Marzke R.O., da Costa L.N., Pellegrini P.S., Willmer C.N.A., Geller
M.J., 1998, \apj, 503, 617

\bibitem[]{Mck02}
McKay T.A., et al, \apj, 2002, 571, L85

\bibitem[]{Mer02}
Merch\'an M., Zandivarez A., 2002, \mnras , 335, 216

\bibitem[]{Mih96}
Mihos J.C., Hernquist L., 1996, \apj , 464, 641

\bibitem[]{Mo96}
Mo H.J., White S.D.M., 1996, \mnras , 282, 347

\bibitem[]{Mo98}
Mo H.J., Mao S., White S.D.M., 1998, \mnras , 295, 319

\bibitem[]{Mo04}
Mo H.J., Yang X., van den Bosch, F.C., Jing Y.P., 2004, \mnras ,
349, 205

\bibitem[]{Mo05}
Mo H.J., Yang X., van den Bosch, F.C., Katz N., 2005, preprint
(astro-ph/0506516)

\bibitem[]{Moo96}
Moore B., Katz N., Lake G., Dressler A., Oemler A., 1996, Nature, 379,
613

\bibitem[]{Moo99}
Moore B., Lake G., Quinn T., Stadel J., 1999, \mnras , 304, 465

\bibitem[]{nav97}
Navarro J.F., Frenk C.S., White S.D.M, 1997, \apj, 490, 493

\bibitem[]{Nor02}
Norberg P., et al., 2002, \mnras , 332, 827

\bibitem[]{Oem74}
Oemler A., 1974, \apj , 194, 1

\bibitem[]{Oka01}
Okamoto T., Nagashima M., 2001, \apj , 547, 109

\bibitem[]{Osm04}
Osmond J.P.F., Ponman T. J., 2004, \mnras, 350, 1511

\bibitem[]{Pog99} 
Poggianti B.M., Smail I., Dressler A., Couch W.J.,
  Barger A.J., Butcher H., Ellis R.S., Oemler A.J., 1999, \apj , 518,
  576

\bibitem[]{Pos84}
Postman M. Geller M.J., 1984, \apj , 281, 95

\bibitem[]{Pos96}
Postman M. et al, 1996, AJ, 111, 615

\bibitem[]{Pra03}
Prada F., Vivitska M., Klypin A., Holtzman J.A., Schlegel D.J., Grebel
E.K., Rix, H.-W., Brinkmann J., McKay T.A., Csabai I., 2003, \apj, 598, 260

\bibitem[]{Ram87}
Ramella M., Giuricin G., Mardirossian F., Mezzetti M., 1987, \aap, 188, 1 

\bibitem[]{Ram89}
Ramella M., Geller M.J., Huchra J.P., 1989, \apj ,344, 57

\bibitem[]{Sch98}
Schlegel D.J., Finkbeiner D.P., Davis M., 1998, \apj , 500, 525

\bibitem[]{She96}
Shectman S.A., Schechter P.L., Oemler A., Tucker D.L., Kirshner R.P.,
Lin H., 1996, \apj , 470, 172

\bibitem[]{SMT01}
Sheth R.K., Mo H.J., Tormen G., 2001, \mnras , 323, 1

\bibitem[]{Spi58}
Spitzer L.Jr., 1958, \apj , 127, 17

\bibitem[]{Spr01}
Springel V., White S.D.M., Tormen G., Kauffmann G., 2001, \mnras , 328, 726

\bibitem[]{Sto02}
Stoughton C., et al., 2002, \aj , 123, 485

\bibitem[]{Str01}
Strateva I., et al., 2001, \apj, 122, 1861

\bibitem[]{Tan04}
Tanaka M., Goto T., Okamura S., Shimasaku K., Brinkman J., 2004, \aj,
128, 2677

\bibitem[]{Tan05}
Tanaka M., Kodama T., Arimoto N., Okamura S., Umetsu K., Shimasaku K.,
Tanaka I., Yamada T., 2005, \mnras, 362, 268

\bibitem[]{Too72}
Toomre A., Toomre J., 1972, \apj , 178, 623

\bibitem[]{Tra01}
Tran K.-V., Simard L., Zabludoff A.I., Mulchaey J.S., 2001,  \apj, 549, 172

\bibitem[]{Tu00}
Tucker D.L., et al., 2000, \apjs , 130, 237

\bibitem[]{vdB99}
van den Bosch F.C., Lewis G.F., Lake G., Stadel J., 1999, \apj , 515, 50

\bibitem[]{BYM03}
van den Bosch F.C., Yang X., Mo H.J., 2003, \mnras , 340, 771

\bibitem[]{vdB04}
van den Bosch F.C., Norberg P., Mo H.J., Yang X.H., 2004, \mnras, 352,
1302

\bibitem[]{BYMN05}
van den Bosch F.C., Yang X.H., Mo H.J, Norberg P., 2005a, \mnras, 356,
1233

\bibitem[]{vdB05}
van den Bosch F.C., Weinmann S.M., Yang X.H., Mo H.J, Li C., Jing Y.P., 2005b, 
\mnras, 361, 1203

\bibitem[]{BTG05c}
van den Bosch F.C., Tormen G., Giocoli C., 2005c, \mnras, 359, 1029

\bibitem[]{Wei05}
Weiner B.J., et al., 2005, \apj , 620, 595

\bibitem[]{Whi02}
White M. \& Kochanek C.S., 2002, \apj, 574, 24

\bibitem[]{Whi95}
Whitmore B.C., 1995, Groups of Galaixes, ASP Conference Series,
Vol. 70, 41

\bibitem[]{Whi93}
Whitmore B.C., Gilmore D.M., Jones C., 1993, \apj, 407, 489

\bibitem[]{Wir83}
Wirth A., 1983, \aj, 274, 541

\bibitem[]{YMB03}
Yang X., Mo H.J., van den Bosch F.C., 2003, \mnras , 339, 1057

\bibitem[]{YMBJ05a}
Yang X., Mo H.J., van den Bosch F.C., Jing Y.P., 2005a, \mnras , 356, 1293 
(YMBJ)

\bibitem[]{YMBJ05b}
Yang X., Mo H.J., van den Bosch F.C., Jing Y.P., 2005b, \mnras , 357, 608

\bibitem[]{YMJB05}
Yang X., Mo H.J., Jing Y.P., van den Bosch F.C., 2005c, \mnras , 358, 217

\bibitem[]{Yan05}
Yang X, Mo H.J., van den Bosch F.C., Weinmann S.M., Li C., Jing Y.P., 2005d, 
\mnras , in press (astro-ph/0504477)

\bibitem[]{Yor00}
York D.G., et al., 2000, \aj , 120, 1579

\bibitem[]{Zab98}
Zabludoff A.I., Mulchaey J.S., 1998, \apj , 496, 39

\bibitem[]{Zab00}
Zabludoff A.I., Mulchaey J.S., 2000, \apj , 539, 136

\bibitem[]{Zeh04}
Zehavi I., et al, 2004, submitted to \apj, astro-ph/0408569

\bibitem[]{Zhe04} 
Zheng Z., et al., 2004, preprint (astro-ph/0408564)

\bibitem[]{Zuc97}
Zucca E., et al., 1997, \aap, 326, 477

\end{thebibliography}
\end{document}